\documentclass[pdflatex,sn-vancouver-num]{sn-jnl}

\usepackage{chemformula} 
\usepackage[T1]{fontenc}
\usepackage[utf8]{inputenc}
\usepackage{graphicx}%
\usepackage{multirow}%
\usepackage{amsmath,amssymb,amsfonts}%
\usepackage{amsthm}%
\usepackage{mathrsfs}%
\usepackage[title]{appendix}%
\usepackage{xcolor}%
\usepackage{textcomp}%
\usepackage{manyfoot}%
\usepackage{booktabs}%
\usepackage{algorithm}%
\usepackage{algorithmicx}%
\usepackage{algpseudocode}%
\usepackage{listings}%

\usepackage{amsmath} 
\usepackage{amssymb} 
\usepackage{adjustbox}
\usepackage{rotating}
\usepackage{booktabs}
\usepackage{lscape}
\usepackage{longtable}
\usepackage{tabularx}
\usepackage[table]{xcolor}
\usepackage{array} 
\usepackage{soul}
\usepackage{colortbl} 

\usepackage{geometry}
\usepackage{setspace}
\usepackage{microtype}
\usepackage{makecell}

\usepackage{setspace}

\usepackage{natbib}
\setcitestyle{super,open={},close={}}

\usepackage{graphicx}
\usepackage{float}
\usepackage{caption}

\newfloat{scheme}{htbp}{los}
\floatname{scheme}{Scheme}
\floatname{chart}{Chart}
\newfloat{graph}{htbp}{loh}

\newcolumntype{Y}{>{\raggedright\arraybackslash}X}

\theoremstyle{thmstyleone}%
\theoremstyle{thmstyletwo}%
\theoremstyle{thmstylethree}%
\begin{document}

\onehalfspacing

\title[Article Title]{A Retinomorphic Optical Spiking Neuron for Camouflaged Object Detection}


\author[1]{\fnm{Srilagna} \sur{Sahoo}}\email{srilagna.sahoo@gmail.com}

\author[2]{\fnm{Adwaaiit} \sur{Pande}}\email{adwaaiit.pande@gmail.com}

\author[3]{\fnm{Kartikey} \sur{Thakar}}\email{kartik.iitb.ue@gmail.com}

\author[2]{\fnm{Shubham} \sur{Sahay}}\email{ssahay@iitk.ac.in}

\author*[1]{\fnm{Saurabh} \sur{Lodha}}\email{slodha@ee.iitb.ac.in}

\affil*[1]{\orgdiv{Electrical Engineering}, \orgname{Indian Institute of Technology Bombay}, \orgaddress{\street{Powai}, \city{Mumbai}, \postcode{400076}, \state{Maharashtra}, \country{India}}}

\affil[2]{\orgdiv{Electrical Engineering}, \orgname{Indian Institute of Technology Kanpur}, \orgaddress{ \city{Kanpur}, \postcode{208016}, \state{Uttar Pradesh}, \country{India}}}

\affil[3]{\orgname{Texas Instruments}, \orgaddress{\city{Bengaluru}, \postcode{560093}, \state{Karnataka}, \country{India}}}


\abstract{Advanced vision systems require retinomorphic, energy-efficient spike-based preprocessing of dynamic visual scenes. Here, we demonstrate multiple retinal preprocessing functionalities by leveraging a Hodgkin-Huxley-based optical spiking neuron (OSHN) that incorporates a two-dimensional anti-ambipolar phototransistor operated in the subthreshold regime to minimize power consumption. OSHN exhibits wavelength- and intensity-sensitive spike encoding with energy consumption per spike of 0.9 pJ under dark, 2 pJ at 480 nm (mid wavelength, M), and 24.5 pJ at 800 nm (long wavelength, L). The low (biological)-to-high spiking rate (0 - 2 kHz) with substantially faster response times (4.2 \textmu s - 1.25 ms) than the human retina (30 ms - 60 ms), reveal OSHN’s fast decision-making capability. OSHN facilitates concurrent spectral-spatial processing by emulating retinal antagonistic center-surround receptive fields (CSRFs) at a single wavelength (480 nm or 800 nm) with varying intensities, visual adaptation (at 480 nm) to prevent system saturation, and L-M cone opponency in midget ganglion cells. Finally, a CSRF-augmented spiking neural network (SNN) has been developed for camouflaged object detection, achieving 4.4$\%$, 10.4$\%$, and 28.4$\%$ improvements in accuracy over conventional SNN on FMNIST, COD10K, and synthetic camouflaged datasets, outperforming existing photoactive spiking architectures while enabling event-driven intelligent edge vision systems.}


\maketitle


Surging demand for advanced visual perception systems across next-generation intelligent applications, such as opto-bionics, autonomous navigation, and edge artificial intelligence (AI), has highlighted the limitations of high latency, limited dynamic range, high power consumption, and data redundancy in conventional vision systems. Their von Neumann architecture requires analog-to-digital conversion circuits to convert sensory analog input signals to digital format, leading to area- and energy-inefficiencies. \cite{Zhou2020} In contrast, the human retina encodes continuous photonic stimuli into voltage spike trains and sends them to the visual cortex for processing, enabling a compact, low-power biological vision system. \cite{Radhakrishnan2021, Wang2023} Bio-inspired neuromorphic vision systems that closely mirror the retina's in-sensor computing can alleviate the von Neumann bottleneck and enable event-driven spike encoding, rather than entire image frames, thereby reducing data redundancy while supporting low-power computation. \cite{npjWang2024, Fu2026, Sahoo2025} Two-dimensional (2D) material-based optoelectronic devices offer a compelling platform for low-power, neuromorphic opto-sensory computing due to their ultra-thin scalability, intrinsic mechanical flexibility, bandgap tunability, and strong light-matter interactions. \cite{Ko2025, Thakar2023, Sahoo2025, Chakrabarty2025, Jawa2022, Thakar2021, Varghese2020}

 Forward phototransduction in the human retina occurs through photoreceptor cells (PCs) $\rightarrow$ to $\rightarrow$ bipolar cells (BCs) $\rightarrow$ to $\rightarrow$ retinal ganglion cells (RGCs), along with two interneurons: horizontal cells (HCs) and amacrine cells (ACs), which are responsible for lateral interactions. \cite{Clifford2002} The HCs collect responses from different PCs and send lateral-inhibitory feedback to the PC-BC synapses, shaping the center-surround receptive field (CSRF) \cite{diamond2017inhibitory} of the BCs and the RGCs, facilitating spatial contrast enhancement \cite{Yun2016}, color opponency \cite{Thoreson2012}, and edge detection \cite{Kim2025} preprocessing functionalities. Other retinal preprocessing functionalities include wavelength and intensity encoding \cite{Laswick2024} and visual adaptation (VA). \cite{Wang2023, Gao2023, Li2024} Biomimicking the VA functionality of PCs has been widely reported using a phototransistor or an optical synaptic architecture within an artificial neural network (ANN)-based machine vision system. \cite{Han2026, Liao2022, Gao2023, Li2024} A step further, ambipolar-WSe$_2$-based phototransistors exhibiting dual-polarity photoresponses have been exploited for replicating graded potential photoresponse characteristics of PCs and BCs, thereby demonstrating simulation of the BC receptive field. \cite{Wang2020, Zhang2022, Han2024} These retinomorphic vision sensory systems have focused mostly on emulating only PC or PC$\rightarrow$BC characteristics, representing a partial emulation of the human retinal path, and processing of continuous-time analog signals. The absence of spike information encoding has restricted their integration into biologically plausible spiking neural network (SNN)-based vision systems.
 
 Few research groups have demonstrated introductory light-to-spike encoding based on a leaky integrate-and-fire (LIF) neuron model utilizing separate blocks of a photoelectric synapse and an MoS$_2$-threshold switch memristor \cite{Pei2021} or an MoS$_2$-based 1T-1C device. \cite{LiH2024} Furthermore, a monolayer MoS$_2$ device has been reported to emulate LIF neuron behavior, exhibiting wavelength-dependent spike responses in an SNN. \cite{Aung2025} Stochastic optical spike encoding in an SNN architecture has also been demonstrated using a combination of a Si-photodiode and a MoS$_2$ FET \cite{Radhakrishnan2021} or a b-AsP/MoTe$_2$ \cite{Wang2023} heterostructure. Although the above studies demonstrate retina-like light-spike encoding, they fail to capture the many preprocessing functionalities that enable the human eye to distinguish minute feature details, such as spatial background contrast relative to the foreground, temporal resolution, and a wide dynamic range. Despite their simplicity and computational efficiency, the IF or LIF neuron models fall short of the biological realism of the Hodgkin-Huxley (HH) model, as they fail to capture the dynamics of ion-channels and the precise shape of the action potential ($V_M$); rendering them unsuitable for accurate and detailed biophysical integration. Hence, it is vital to develop an HH neuronal model-based optical neurocomputing architecture, that can support both biomedical and robotic integration.

This work presents an anti-ambipolar phototransistor based on a single 2D photoactive channel material, WSe$_2$ \cite{Ghosh2022, Thakar2021}, featuring three back gates: a channel gate ($G_C$) and two barrier gates ($G_B$s). The consistent subthreshold operation of the 2D anti-ambipolar phototransistor (2D subthreshold phototransistor, 2DSpT) is enabled by effective electrostatic control through complementary bias sweeping of $G_C$ and $G_Bs$, resulting in low-power operation. The optoelectronically modulated, bell-shaped transfer curve of this anti-ambipolar phototransistor resembles the dynamics of sodium ion (Na$^+$) conductance in biological neurons. \cite{Thakar2023, Romanova2022} An optical spiking neuron based on the HH neuron model (OSHN) has been developed that incorporates the WSe$_2$ phototransistor's (as a Na-channel) behavioral model, implemented in 45 nm CMOS technology, to mimic wavelength-intensity-sensitive (WIS) spike encoding under dark and illumination (480 nm and 800 nm). The WIS spike encoding was achieved with energy consumption per spike (EPS) of 0.9 pJ in the dark and 2 pJ and 24.5 pJ under 480 nm (M, mid wavelength) and 800 nm (L, long wavelength) illumination, respectively. It exhibits a low (biological)-to-high spiking rate (0 - 2 kHz) while achieving response times (4.2 \textmu s  - 1.25 ms) substantially faster than the retina (30 ms - 60 ms). \cite{Masland2017} OSHN outperforms a similar HH-vision system (0.5 Hz - 25 Hz spiking rate and 1.1\texttimes10\textsuperscript{6} pJ EPS at 580 nm) based on Si-photodiodes integrated with a bilayer organic transistor \cite{Laswick2024} by exhibiting 80$\times$ higher spiking rate and 10\textsuperscript{6}-fold lower EPS. Relative to OSHN, a visual neural encoder \cite{TTFSLi2024} exhibiting a higher spiking rate (0.35 MHz – 1.85 MHz) and a comparable visual cue response time (1 \textmu s – 13 \textmu s) also requires 500-fold greater energy per spike (EPS, 1 nJ). The above reported optical spike encoding systems show higher spike rates, but at the cost of higher energy consumption, and relatively few studies have focused on first-spike arrival time, which is an important parameter for quick decision-making capability. These reports exhibit limited retinal functionalities (WIS and VA are widely discussed, whereas RF is rarely addressed) with partial emulation of the retinal path at a single wavelength or under white light with a narrower intensity range, constraining their applicability in advanced vision tasks such as camouflage detection for autonomous vehicles, medical imaging, and military reconnaissance. 

Here, we report a complete PC$\rightarrow$BC$\rightarrow$RGC retinal forward path, along with the lateral-inhibitory interaction from HCs that shapes the CSRF of BCs and RGC spiking responses, has been biomimicked. The CSRF vision helps enhance spatial contrast and emphasize edges, enabling the capture of finer detail at the pixel level. The dual-band detection of 2D-WSe$_2$ enables OSHN to mimic midget RGCs' simultaneous red-green (L-M-cone) chromatic opponency and contrast-resolving capability, which has not yet been addressed, thereby improving sensitivity to both spectral and non-chromatic features for the identification of complex camouflaged scenes at a near and higher (0 - 970 Hz) rate than the retina (0 - 500 Hz). Moreover, VA dependent on CSRF, which includes photopic and scotopic vision, has been demonstrated under 480 nm illumination, showing comprehensive light-evoked calcium-mediated regulation of RGC spiking sensitivity without response saturation within a single framework. Finally, the CSRF-mediated spiking response of OSHN is shown to enable functional demonstration of camouflaged object detection (COD) to assess system-level performance. Conventional SNN encoders directly convert pixel intensities into spike probabilities, leading to low contrast boundaries that are indistinguishable in the temporal spike domain.\cite{fang2021incorporating} Here, the SNN framework has been trained on three different datasets: i) Fashion MNIST (FMNIST, foreground and background are clearly distinct), ii) COD10K-v3 (low contrast for camouflage detection), and iii) a synthetic camouflaged dataset (lower contrast than COD10K-v3 for more challenging camouflage detection), and the test accuracies obtained were 68.9$\%$, 74.9$\%$, and 95.9$\%$, respectively. The CSRF-feature augmented SNN exhibits substantial improvements in classification accuracies of 4.4$\%$, 10.4$\%$, and 28.4$\%$ for the FMNIST, COD10K, and synthetic camouflaged datasets, respectively, compared to a conventional SNN. This validates the developed SNN architecture for low contrast visual scenes, highlighting its potential for energy- and area-efficient retina-like real-time intelligent vision systems surpassing previously reported photoactive spiking architectures.



\newpage
\section{Device Architecture}
A schematic of the 2DSpT comprising three electron beam (e-beam) lithography patterned bottom gates (one $G_C$ and two $G_B$s shorted together, Ti (4 nm)/Au (30 nm)), a 2D dielectric (hBN, 63 nm) layer, and a trilayer (2.2 nm) 2D semiconductor (WSe$_2$) channel on top is presented in Fig.~\ref{arch}a. Subsequently, the source/drain (S/D) electrodes were e-beam patterned and metallized (Cr (4 nm)/Pt (30 nm)/Au (70 nm)) using sputtering. A pseudo-color optical micrograph of the fabricated device is presented in Fig.~\ref{arch}b. Thickness profiles of hBN and WSe$_2$ flakes, characterized by atomic force microscopy (AFM), are shown in Fig.~\ref{arch}c. The Raman scattering peaks at 250.6 cm$^{-1}$ and 257.7 cm$^{-1}$ (Supporting Information, Fig. S1a) are attributed to the E$_{2g}$/A$_{1g}$ phonon modes and the double resonance 2LA(M) mode of WSe$_2$, respectively. \cite{Tonndorf2013} The micro-photoluminescence (PL) spectrum of WSe$_2$, indicating an indirect (I) transition along with two direct transition related peaks (prominent-A and weak-B) at higher energies, is shown in Fig. S1b. \cite{Zeng2013}

\section{Parallels between Biological Retina and Opto-neural System}

\begin{figure} [H]	
{\includegraphics[width=1\textwidth]{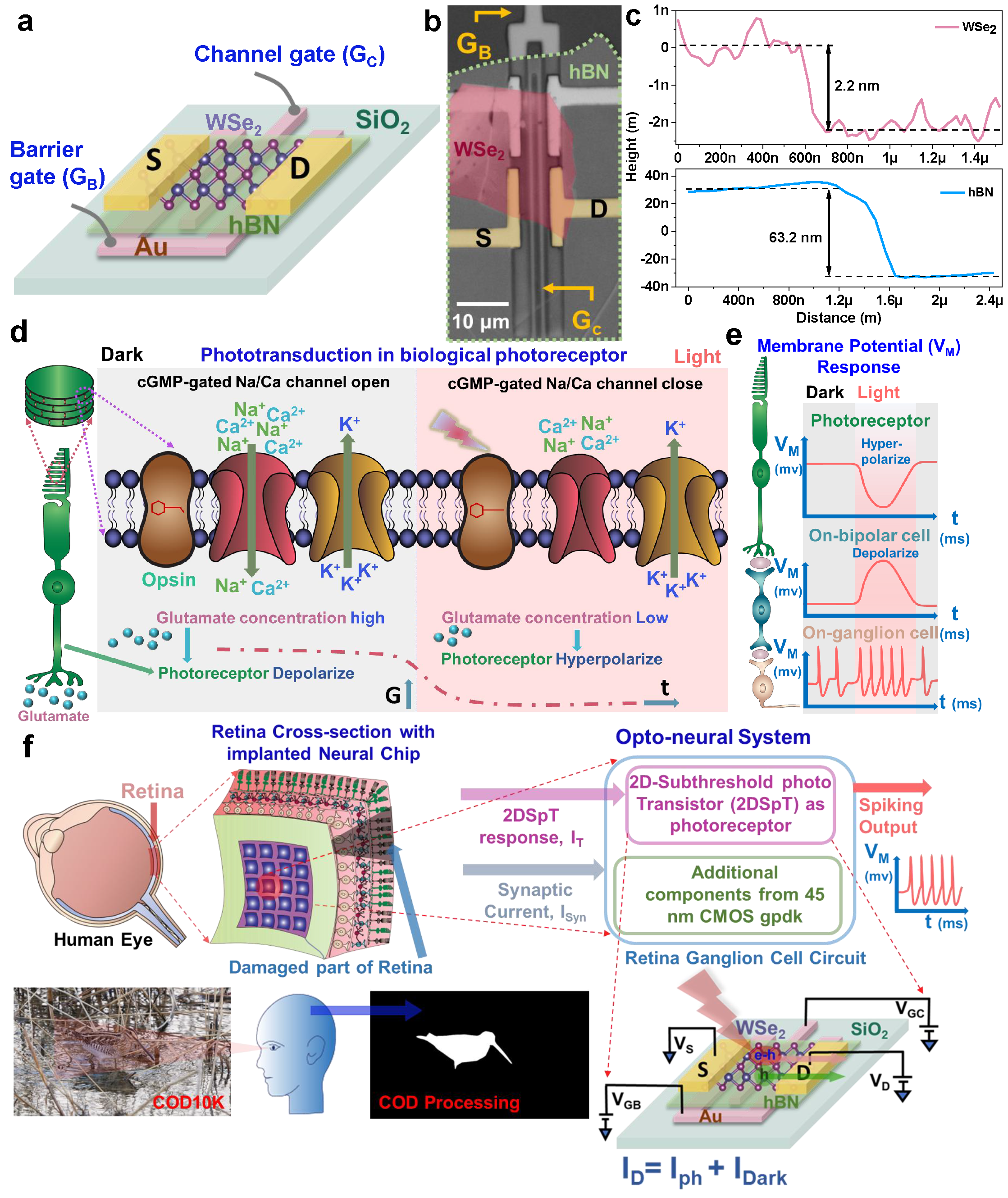}}
\caption{\textbf{Device architecture, the biological phototransduction process, temporal membrane potentials ($V_M$), and artificial retinomorphic vision system.} (a) Device schematic of a three-back-gated (channel gate ($G_C$) and two barrier gates ($G_B$s) shorted together) 2D-WSe$_2$ subthreshold phototransistor (2DSpT) with hBN as the gate dielectric. (b) A false-colored optical micrograph image of the fabricated 2DSpT. The green dashed line indicates the boundary of the hBN flake, and the red shaded region defines the WSe$_2$ flake. (c) Thickness profiles of WSe$_2$ and hBN flakes. (d) The biological phototransduction process, under dark and illuminated conditions, is highlighted in the plasma membrane of the cone photoreceptor cell. This portrays that a decrease in glutamate neurotransmitter concentration leads to a decrease in conductance under light stimulus. (e) The membrane potential responses of biological PC, BC, and RGC. (f) Camouflaged object detection utilizing an opto-neural circuit that integrates a 2DSpT as a photoreceptor to perform light-to-spike encoding while emulating RGC functionality, enabling downstream information processing compatible with the visual cortex.}
\label{arch}
\end{figure}

Phototransduction in retinal PCs involves the absorption of photons by visual pigments (opsins, membrane proteins) present in the membranous disks of human PCs, followed by their conversion into electrical signals through coordinated biochemical and ionic mechanisms. A magnified version of the PC's plasma membrane consisting of opsin along with the cyclic guanosine 3$'$, 5$'$-monophosphate (cGMP) gated Na$^+$/Ca$^{2+}$ \cite{cGMPLi2023} and potassium (K$^+$) channels is shown in Fig.~\ref{arch}d. 

Under dark, the cGMP-gated channels are open, leading to a large inward Na$^+$/Ca$^{2+}$ current, which is balanced by a K$^+$ outward current, causing the PC to depolarize.\cite{Molday1998} In this condition, glutamate (a neurotransmitter) concentration is high at the PC-BC synapse. Under illumination, initiation of the G-protein signaling cascade leads to closure of cGMP-channels, reducing the inward Na$^+$/Ca$^{2+}$ current.\cite{Klapper2016} Now, the membrane potential is dominated by K$^+$ outward current, which hyperpolarizes the membrane. This results in a decrease in glutamate concentration (decrease in membrane conductance, G) at the PC-BC synapse compared to the dark. \cite{Boccuni2022} The K-channels require depolarization to remain open; hyperpolarization progressively deactivates them, facilitating recovery from the light stimulus and thereby shaping the temporal photoresponse of the PCs. The graded potential photoresponse of a PC is shown in Fig.~\ref{arch}e. The light-invoked glutamate concentration variation also triggers a BC response, resulting in either hyperpolarization or depolarization and a graded temporal potential response. The only retinal neuron that generates a spiking response (action potential, $V_M$) is the RGC, as shown in Fig.~\ref{arch}e. \cite{Awatramani2000, Ichinose2022}

The degeneration of retinal PCs due to diseases such as retinitis pigmentosa and age-related macular degeneration, or partial retinal damage, prevents the conversion of light into spikes, resulting in visual dysfunction. \cite{Mills2017, Wu2023} Artificial vision systems offer a potential solution for restoring vision through the development of emerging opto-neural technologies. Here, we demonstrate camouflaged object detection utilizing an opto-neural system that imitates the retinal preprocessing functionalities implemented using an RGC spiking circuit (Fig.~\ref{arch}f) by incorporating 2DSpT as a PC and additional circuit components from a 45 nm CMOS general process design kit (gpdk) for the realization of BC$\rightarrow$RGC connections, along with the HC interneuron.


\section{\textbf{Optoelectronic Characterization and Device Physics}}

\begin{figure} [H]	
{\includegraphics[width=1\textwidth]{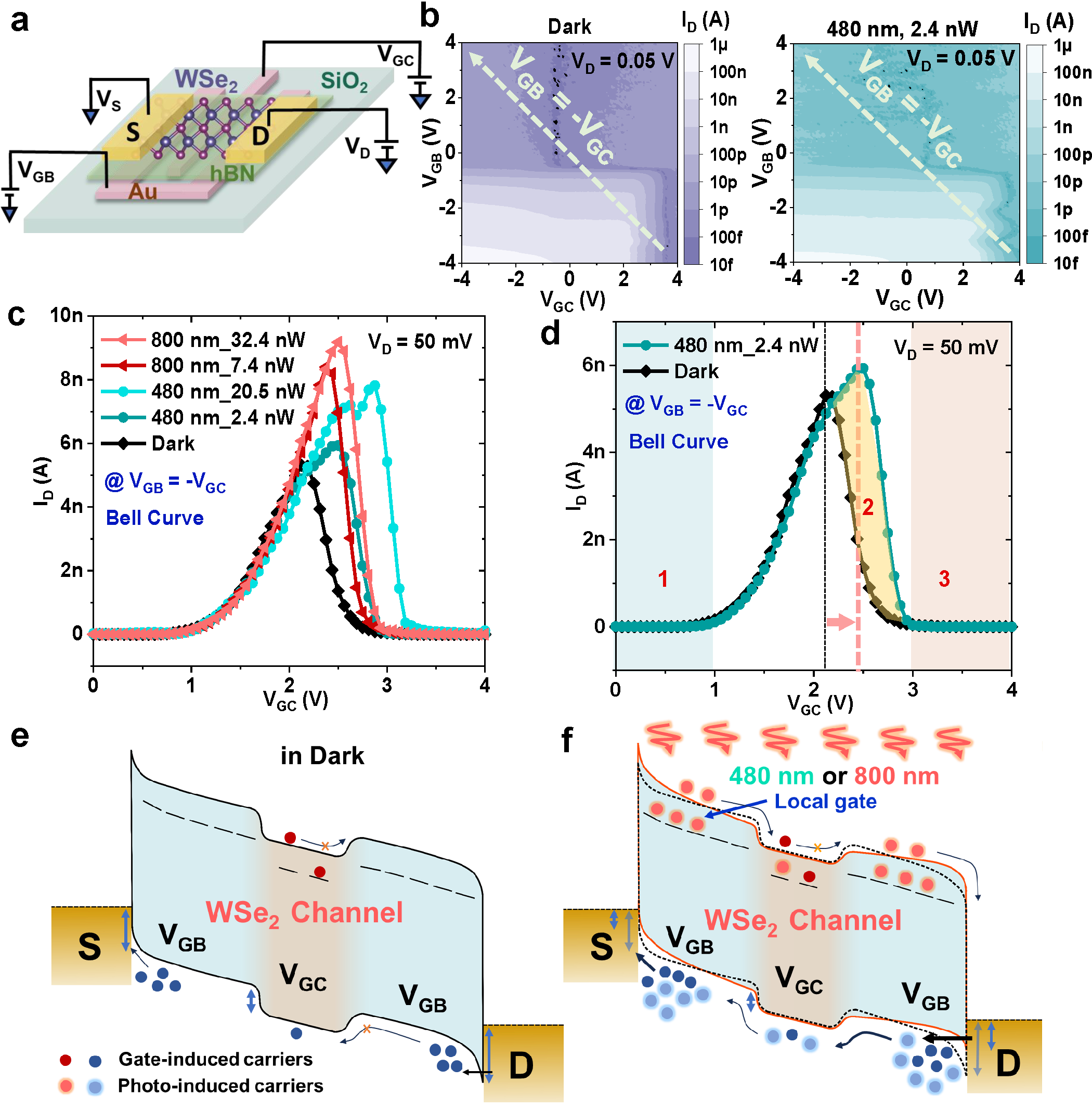}}
\caption{\textbf{Optoelectronic anti-ambipolar response (bell-shaped transfer characteristic) of 2DSpT and underlying device physics.} (a) Biasing schematic for 2DSpT is configured to measure the device drain current ($I_D$) as a function of the channel gate ($V_{GC}$) and barrier gate ($V_{GB}$) biases. The two color maps show 2DSpT drain currents (b) in dark and 480 nm (2.4 nW) illuminated conditions with respect to $V_{GC}$ and $V_{GB}$. (c) Bell-shaped $I_D$ was extracted along the diagonal arrow marked in the color maps under dark, 480 nm and 800 nm illuminated conditions for varying intensities, where $V_{GB}$ = -$V_{GC}$. (d) Right-shifted bell-shaped current at 480 nm (2.4 nW) relative to dark condition is shown to elucidate the device physics under (e) dark and (f) 480 nm or 800 nm illumination (marked as regime 2). Increase of $V_{GC}$ from 0 V to 4 V causes a continuous increase in the channel barrier (CB), and decrease of $V_{GB}$ from 0 V to -4 V causes a reduction in the Schottky barrier (SB). Hence, the current is limited by SB in regime 1 and by CB in regime 3. Cyan and peach color shades in regimes 1 and 3 are added according to the dominance of SB and CB, respectively.
}
\label{elec}
\end{figure}

The schematic shown in Fig.~\ref{elec}a illustrates the biasing scheme of 2DSpT for optoelectronic characterization. The tri-back-gated structure exhibits a p-FET transfer characteristic (drain current vs. channel gate bias, $I_D$-$V_{GC}$) through the concurrent sweep of $V_{GC}$ and $V_{GB}$ ($V_{GC}$ = $V_{GB}$), shown in Supporting Information, Fig. S2a. The transfer characteristics switched to a bell-shaped profile (Supporting Information, Fig. S2b) when the channel gate and barrier gate biases were swept in opposite directions ($V_{GC}$ = -$V_{GB}$). The transconductance (Supporting Information, Fig. S2c), $g_m$, extracted from Fig. S2b, shows both positive and negative $g_m$ profiles. The transition from the negative $g_m$ to positive $g_m$ regime with respect to $V_{GC}$ plays a key role in setting up a regenerative positive feedback loop with enhancement in drain current till a peak is attained. Further decline in $V_{GC}$ suppresses the drain current, introducing self-inactivation and resetting the device. \cite{Beck2020, Thakar2023} The device currents, $I_D$, under dark and 480 nm (2.4 nW) illumination as a function of $V_{GB}$ and $V_{GC}$ are shown in Fig.~\ref{elec}b. The other color maps of device current under 480 nm (20.5 nW) and 800 nm (7.4 nW, 32.4 nW) are provided in Supporting Information, Fig. S2d-f. Bell-shaped $I_D$ loci were extracted along $V_{GB}$ = -$V_{GC}$ (white arrows in Fig.~\ref{elec}b) from the above-mentioned color maps and shown in Fig.~\ref{elec}c. Under illuminated conditions (480 nm or 800 nm), the bell-curve peak current is enhanced and right-shifted with increasing illumination power compared to dark. $g_m$ profiles extracted from Fig.~\ref{elec}c are shown in Supporting Information, Fig. S2g.

The right-shifted bell curve (Fig.~\ref{elec}d) can be explained by dividing the entire range of $V_{GC}$ sweep into 3 regimes. In regimes 1 ($V_{GC}$ < 1 V) and 3 ($V_{GC}$ > 3 V), the illuminated currents are suppressed like the dark current. The current is suppressed by barrier gate bias, $V_{GB}$, in regime 1 due to limited hole injection at the drain-channel Schottky barrier (SB), and in regime 3, due to an increase in channel barrier (CB) with $V_{GC}$. \cite{Thakar2023} Regime 2, the yellow shaded part where the illuminated bell curve is distinct from the dark one, can be explained by the energy band diagrams (EBDs) shown in Fig.~\ref{elec}e-f and Supporting Information 3. The WSe$_2$-2DSpT exhibits dominant p-FET behavior, with holes as majority carriers compared to electrons for the entire sweep range of $V_{GC}$ and $V_{GB}$, Supporting Information 2. At $V_{GC}$ = 2.5 V and $V_{GB}$ = -2.5 V, the EBD across the $G_B$/hBN/WSe$_2$ stack shows hole accumulation in contrast to hole depletion across the $G_C$/hBN/WSe$_2$ stack at the hBN/WSe$_2$ interface (Supporting Information 3). The S (D) SB favors hole collection (injection) and the CB blocks the carrier transport in the dark state (Fig.~\ref{elec}e). Under illumination, uniform generation of electron-hole pairs across the channel results in an increase in hole accumulation at the hBN/WSe$_2$ interface across the $G_B$/hBN/WSe$_2$ stack, which leads to a decrease in hole S/D SBs (Fig.~\ref{elec}f) and an increase in hole current. However, the photogenerated electrons are trapped in the dielectric across the channel, acting as a local gate, leading to a positive shift in the threshold voltage. Thus, we obtain an enhanced, right-shifted illuminated net current.


\section{\textbf{Neuromorphic Emulation of Retinal Preprocessing Functionalities}}

\subsection{\textbf{Wavelength- and Intensity-Sensitive Encoding}}

Wavelength- and intensity-sensitive spike encoding forms the foundation of neuromorphic vision systems, facilitating retina-like event-driven per-pixel processing (akin to cones' wavelength discrimination and rods' intensity sensing) with high dynamic range perception. \cite{Risi2020} A retina-inspired optical spiking neuron based on the HH model has been implemented that operates by leveraging photo-modulated anti-ambipolarity of 2DSpT (Supporting Information, Fig. S4a). 2DSpT mimics the Na-channel current of the HH neuron, while K-channel, cell membrane capacitance ($C_M$), and leakage conductance ($g_L$) pathways are realized using additional components from a 45 nm CMOS gpdk (Supporting Information, Table S1). The two input signals that modulate OSHN's spiking response are the $I_{syn}$ and the Enable (E) signal. $I_{syn}$ is the presynaptic current received at the dendritic end, which decides the excitability of the neuron, and E is a control signal that controls the dark and illumination state response of 2DSpT during circuit operation. Comprehensive details on 2DSpT-OSHN integration and circuit operation are presented in Supporting Information 4. 

\begin{figure} [H]	
{\includegraphics[width=1\textwidth]{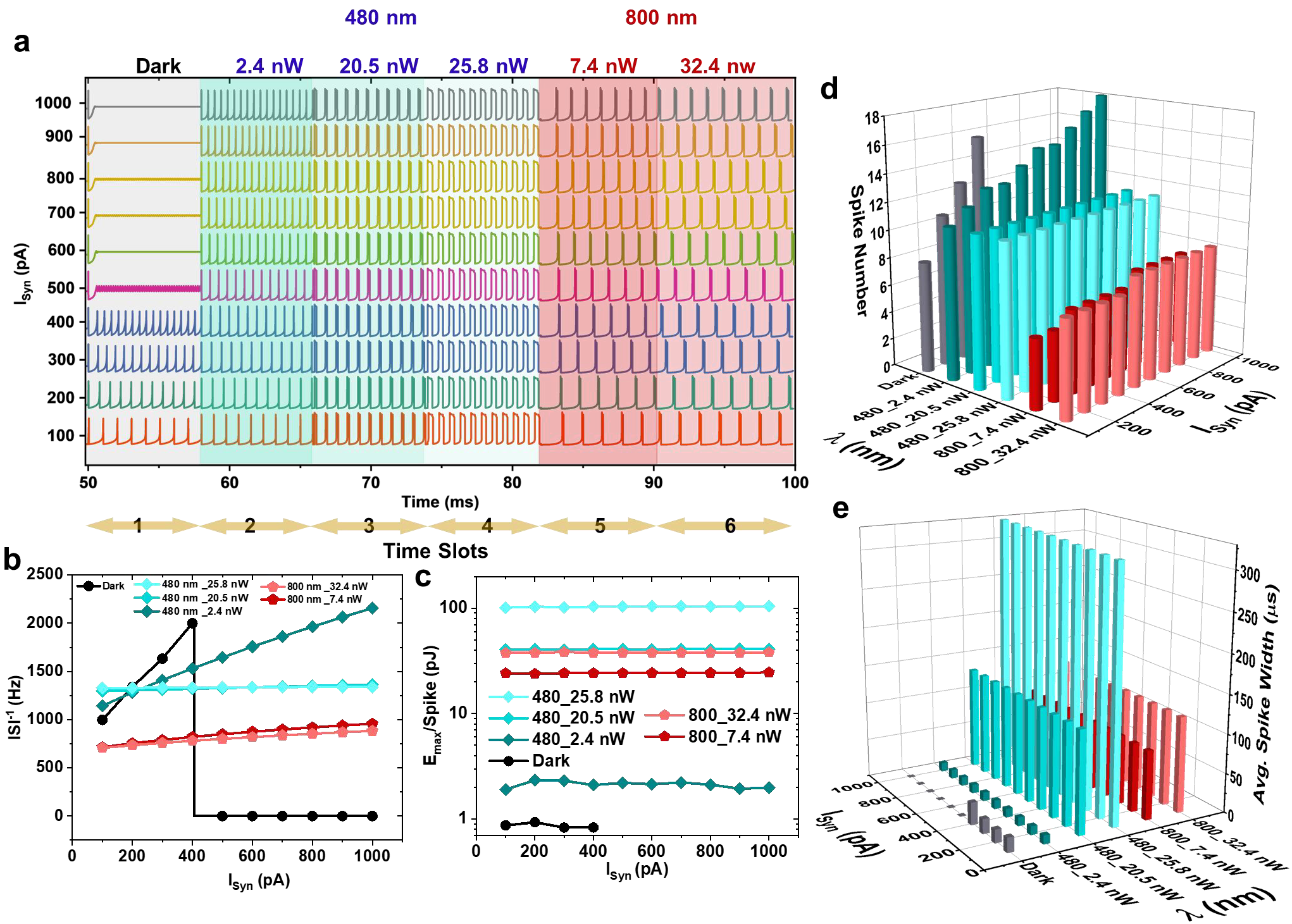}}
\caption{\textbf{Wavelength- and intensity-sensitive spike encoding by a biomimetic optical spiking HH-neuron (OSHN) incorporating 2DSpT as a photo-modulated Na-channel and the spiking properties under dark and 480 nm, and 800 nm illumination at varying intensities.} (a) Action potential, $V_M$ of OSHN as a function of external synaptic current, $I_{syn}$. Time slots have been numerically assigned based on changes in illumination conditions. (b) The spike rate, $ISI^{-1}$, and (c) the maximum energy consumption per spike (EPS) extracted from (a) as a function of $I_{syn}$. Additional spike properties, including (d) spike number and (e) average spike width vs. $I_{syn}$, as extracted from (a).}
\label{ckt}
\end{figure}

Leveraging the photo-modulated anti-ambipolarity of 2DSpT, WIS optical neural spiking is demonstrated (Fig. 3a), similar to that of RGCs. A mid-visible (M) wavelength (480 nm) and a long near-IR (L) wavelength (800 nm) have been chosen for illustration. The inclusion of NIR encoding extends the range of applications such as low-light surveillance, defense, biomedical imaging, and material- and structural-specific analysis. \cite{Zhu2023, Qiao2022} Fig.~\ref{ckt}a illustrates OSHN action potential as a function of incremental presynaptic current ($I_{syn}$: 100 pA - 1000 pA) under dark (time slot 1), 480 nm (for 3 different powers: 2.4 nW, 20.5 nW, and 25.8 nW, time slots 2-4), and 800 nm (for 2 different powers: 7.4 nW and 32.4 nW, time slots 5 and 6). In dark state, OSHN continues to spike till $I_{syn}$ = 400 pA, and beyond this, it stops spiking due to excess depolarization. \cite{HHWang2023, Lee2024} This could be attributed to the rapid rise in $V_M$ at high $I_{syn}$ and its settling at a higher potential than the resting potential without recovery (Supporting Information, Fig. S5a). This prevents 2DSpT from entering a negative $g_m$ regime due to inactivation of Na-regenerative feedback (Fig. S5b), thereby suppressing spike initiation. However, under 480 nm and 800 nm illumination, OSHN does not exhibit such a depolarization block but instead shows spikes across the entire $I_{syn}$ range. 

The spike rate ($ISI^{-1}$: inverse of inter-spike interval, Fig. S6a) extracted from Fig.~\ref{ckt}a for all the illumination conditions is shown in Fig.~\ref{ckt}b. $ISI^{-1}$ increases monotonically as a function of $I_{syn}$, with a steeper slope in dark than under illumination. This is likely due to the negative-to-positive $g_m$ transition with respect to $V_{GC}$ being better aligned with the dark-state current of 2DSpT. Under 480 nm or 800 nm illumination, as intensity increases, the slope flattens. At lower intensities, a negative $g_m$ is repeatedly attained, leading to regular spiking; at higher intensities, $V_m$ is overdriven, pushing $V_{GC}$ far beyond the bell-shaped curve, preventing the circuit from entering regenerative feedback. Hence, as in biological photoreceptors, the 2DSpT only responds to significant changes in contrast (time slots 1-2 and 4-5, due to wavelength changes) until it receives sufficient illumination (time slots 2-3, due to intensity change). At higher illumination, the spiking rate of OSHN saturates (time slots 3-4 and 5-6 due to the intensity change), showing reduced gain. It responds only to contrast rather than absolute intensity. The change in $ISI^{-1}$ with respect to $I_{syn}$ becomes quasi-constant. This property resembles visual adaptation in biological vision.

The maximum energy per spike (EPS) versus $I_{syn}$ for the entire circuit has been evaluated (Supporting Information, Fig. S6a) to be as low as 0.9 pJ for dark, 2 pJ for 480 nm (2.4 nW), and 24.5 pJ for 800 nm (7.4 nW), as shown in Fig.~\ref{ckt}c. EPS increases with increasing intensity due to the number of spikes remaining almost constant with increasing $I_{syn}$ (Fig.~\ref{ckt}d) accompanied with an increase in spike width (Fig.~\ref{ckt}e). The average time to the first spike (TFS) (Supporting Information, Fig. S6b) after stimulus onset is as low as 4.5 \textmu s  (Fig. S6c), which is much faster than the human retina, whose response time is as high as 30 ms to 60 ms. \cite{Masland2017} Hence, an energy-efficient, biorealistic OSHN with response time much faster than the retina with WIS spike encoding is demonstrated.


\newpage
\subsection{\textbf{Visual Adaptation: Photopic and Scotopic Vision}}

OSHN's light-adaptive vision capability, a core preprocessing function preceding spiking transmission to the visual cortex, was also examined. Visual adaptation enhances weak-signal sensitivity by improving the signal-to-noise ratio while attenuating responses to strong inputs to prevent saturation. This dynamic regulation of retinal sensitivity across a 10000-million-fold range of environmental illumination (from starlight to sunlight) is exhibited by rods and cones. \cite{Barbur2010, Reeves2009} Rods, more intensity-sensitive, dominate low-light vision (scotopic vision), while cones govern daylight and color vision (photopic vision). \cite{Hisatomi2002, Jacobs2002} This adaptability arises from rapid pupillary constriction, photopigment bleaching, illumination-dependent rod–cone dominance, and calcium-regulated phototransduction cascades.\cite{Barbur2010, Nikolaeva2023, Reeves2009}

Scotopic vision involves slow recovery of visual sensitivity upon returning to low light after prolonged exposure to intense illumination. Whereas photopic vision rapidly adapts by reducing the visual sensitivity during dark-to-a minimum detected brightness transition, the rods saturate and shift visual processing to cones, leading to transient eye dazzle and then gradual recovery of vision (Fig.~\ref{adap}a). \cite{Demb2008, Lamb2010} Weber's law states that the just noticeable difference (JND, $\Delta$I) in two intensities is a constant proportion of initial light intensity (I): $\frac{\Delta I}{I} = k$. \cite{Rushton1963, Reeves2009, Liao2022} This indicates a proportional increase in visual threshold with illumination, enabling contrast encoding rather than absolute intensity, reflected in 2DSpT photoresponse that exhibits a right-shifted bell curve with increasing incident photons (Fig.~\ref{elec}c).

Calcium ions (Ca$^{2+}$) play a key role in VA by regulating the sensitivity or gain of the retinal forward path. Light-evoked Ca$^{2+}$ concentration drop yields a graded photoresponse at postsynaptic terminals of PCs and BCs\cite{Krizaj2002}, whereas the Ca$^{2+}$-mediated large (BKCa) and small (SKCa) conductance K-channels in RGCs effectively shape the spike and support $V_M$ frequency adaptation by controlling post-hyperpolarization times and synaptic currents. During low (high) light adaptation, BKCa-channels increase (decrease) excitatory postsynaptic currents (EPSCs), thereby adjusting RGC gain and facilitating light-evoked bursting spike patterns. \cite{ZHONG2013, Demb2008} Hence, the OSHN circuit has been modified ($D_0$-$C_{Ca}$ is a leaky integrator) to demonstrate Ca$^{2+}$-mediated spikes (Supporting Information, Fig. S4b). The presynaptic current received at dendrite end of the ganglion cell is shaped by excitatory glutamate (released from BC), inhibitory $\gamma$-aminobutyric acid (GABA), and glycine neurotransmitters released from ACs \cite{Lagnado1998} and can be modeled by a double exponential \cite{Hartveit1999, Awatramani2000} function:
{\setlength{\abovedisplayskip}{0.5cm}
\setlength{\belowdisplayskip}{0.5cm}
\begin{equation}
\text{$I_{syn}$}(t) = A \left(1 - \exp\left(-\frac{t - t_{\text{onset}}}{\tau_{\text{r}}}\right)\right)
\exp\left(-\frac{t - t_{\text{onset}}}{\tau_{\text{f}}}\right), \quad t \ge t_{\text{onset}}
\end{equation} 
where $\tau_r$ = rising time constant (20 ms), $\tau_f$ = falling time constant (20 ms) and $t_{onset}$ = signal arrival time (100 ms).

Mimicking scotopic adaptation, exposure to high intensity of 480 nm (20.5 nW) illumination with low EPSCs, $ISI^{-1}$ is maintained around 500 Hz. A sudden decrease in intensity (20.5 nW to 2.4 nW) causes a transient suppression of spiking activity, similar to the initial moment of invisibility when moving from bright daylight to a dimly lit room. This can be attributed to OSHN failing to elicit a spike due to a low $I_{syn}$ (low sensitivity at a higher range of environmental illumination). Over time, OSHN (Fig.~\ref{adap}b) gradually adapts to the ambient light condition by increasing the spike rate (no spike to 650 Hz) with increased $I_{syn}$ (increasing sensitivity during low illumination), Fig.~\ref{adap}c. The initial rate of increase in spike rate is higher than the later rate, indicating effective sensitivity or gain control, as in BKCa/SKCa-mediated RGC spiking. 

During photopic adaptation of OSHN, when the exposure condition is changed from dark to low intensity of 480 nm (2.4 nW) illumination, the dark spike rate (550 Hz) shows a sudden jump and saturates at 625 Hz, akin to highly sensitive (high $I_{syn}$) rod saturation under bright conditions. The visual path progressively changes to low-sensitive cones (low $I_{syn}$), and the action potential (Fig.~\ref{adap}d) effectively adapts to the prevailing ambient intensity by decreasing the spiking rate (Fig.~\ref{adap}e). In summary, Ca$^{2+}$-mediated VA occurs across multiple layers of the retina, from PC/BC synaptic modulation to controlled-gain RGC spiking. 
  

\begin{figure} [H]	
{\includegraphics[width=1\textwidth]{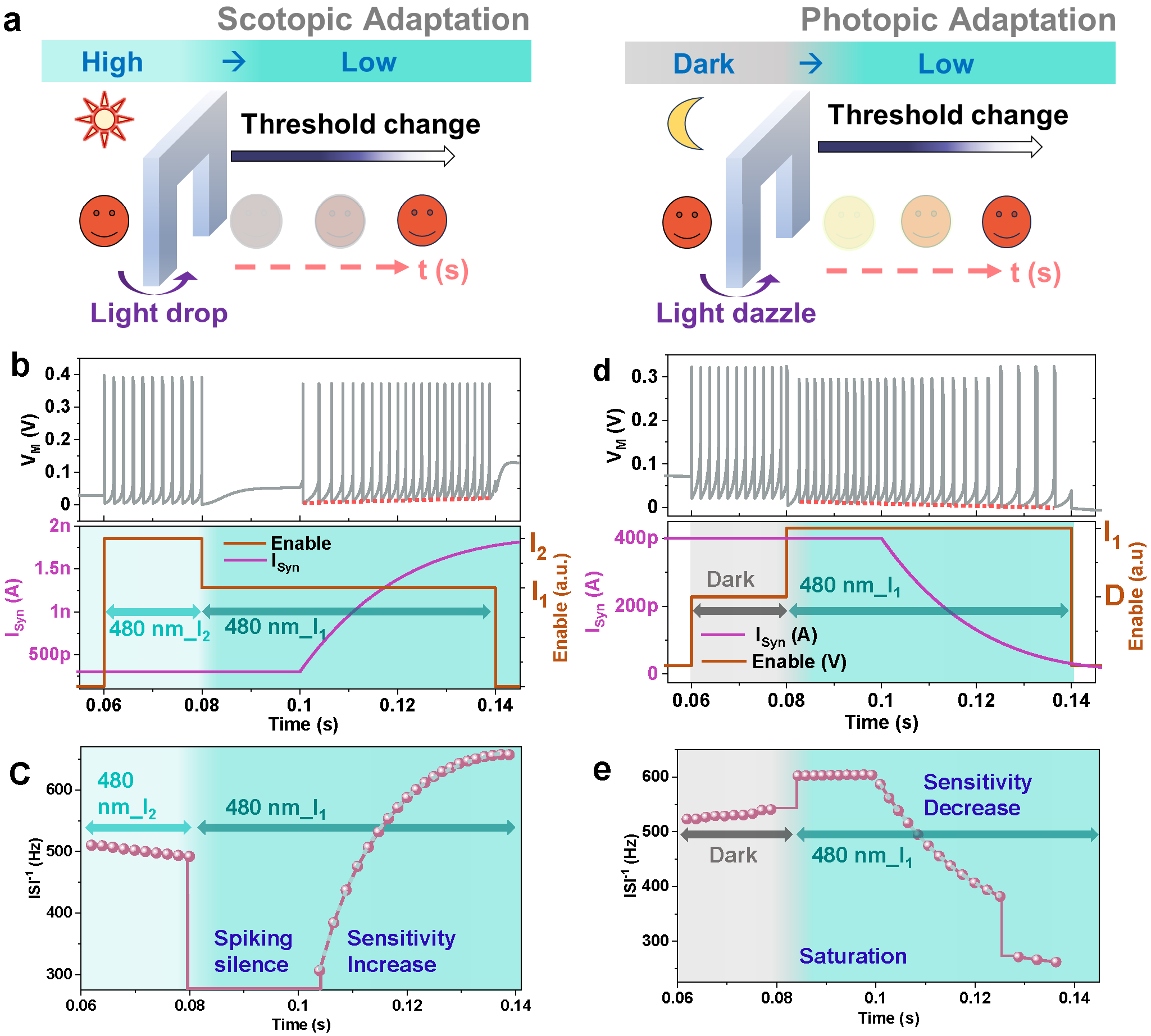}}
\caption{\textbf{Biomimicking of visual adaptation.} (a) The concepts of scotopic and photopic vision illustrate illumination intensity adaptation and changes in detection threshold. Demonstration of (b) scotopic and (d) photopic adaptive action potential ($V_M$) by OSHN that responds to variation in illumination by adjusting the post-hyperpolarization times and spiking threshold of $V_M$, portraying effective gain control of ganglion cells. (c), (e) The extracted instantaneous spiking rate from (b) and (d), respectively, illustrates the effective emulation of sensitivity adjustment of rods and cones in response to variation in illumination.}
\label{adap}
\end{figure}

\subsection{\textbf{Center-Surround Receptive Field-mediated Vision}}
Beyond visual adaptation, the retina serves as an active computational front end, performing sophisticated parallel preprocessing and spike-based encoding of visual cues to amplify spatiotemporal contrast for edge and motion detection. The propagation of light through the layered retinal neurons and interneurons, along with the signal-processing pathways, is depicted in Fig.~\ref{CSRF2}a. A retinal micro-unit is highlighted, consisting of 3 PCs, 1 BC, 1 RGC, and 2 HCs, organized in a CSRF configuration (detailed explanation in Supporting Information 7). Out of the 3 cone cells, one directly communicates (center PC) with the BC$\rightarrow$RGC path (forward path), and the other two PCs (surround PCs) communicate through HCs to form the lateral inhibitory path. Further, BCs are broadly classified into two types, ON-center and OFF-center, depending on the receptors they express that bind glutamates, released by PCs. This defines the receptive field of BCs. During dark state, PC glutamate release is high, and the ON-center BCs hyperpolarize; under light, PC hyperpolarization reduces glutamate release, and bipolar cells send an excitatory signal. The OFF-center bipolar cells are active in the absence of light in the center of their receptive field.\cite{Ichinose2022} The ON-center BCs converge with excitatory input to ON-center RGCs (ON-RGCs), while OFF-center ganglion cells (OFF-RGCs) are excited by OFF-center bipolar cells. \cite{Balasubramanian2009} 

The light-evoked RGC response is much more complicated than ON- and OFF-center, as it processes surrounding information via HCs (which contain multiple surrounding PCs) and ACs (which include numerous surrounding BCs), resulting in a center-surround receptive field. Biological ON-RGC spiking response under varying center-surround illumination contrasts as shown in Fig.~\ref{CSRF2}b, exhibits increased spiking rate during center illumination (2) and suppression during surround illumination (4). Diffuse illumination (3) elicits minimal spiking activity of RGC and near-zero response under dark (1). The OFF-RGC has an opposite response compared to the ON-RGC. \cite{Ichinose2022} To mimic the ON-ganglion cell spiking response, the OSHN circuit (Fig. S4b) has been modified (Fig.~\ref{CSRF2}c), such that it integrates 3 2DSpTs as 3 PCs (the center PC is flanked by two surrounding PCs) whose responses feed directly and indirectly through a horizontal feedback current-steering circuit (Supporting Information, Fig. S7) to a resistor (BC) and further to an HH neuron based ganglion spiking circuit. The circuit components are listed in Supporting Information, Table S2-3. 

The CSRF-dependent RGC spike rates are determined by the contrast between center and surround PCs under dark, 480 nm, and 800 nm illumination. First, OSHN spiking response has been investigated under different CS-light patterns of dark and 480 nm illumination (Fig.~\ref{CSRF2}d). The control signals that modulate the dark and photoresponse of 2DSpT during circuit operation for center-2DSpT (C) and surrounding-2DSpT (S) (Fig.~\ref{CSRF2}c) are Enable-C and Enable-S, respectively. Here, the external $I_{Syn}$ effect is kept silent only to see the CS-light-dependent spiking. The generated spatiotemporal action potential, $V_M$, corresponding to each CS-light pattern is shown in Fig.~\ref{CSRF2}d. The extracted raster plot and $ISI^{-1}$ from the spatiotemporal $V_M$ data are shown in the next two panels of Fig.~\ref{CSRF2}d. The $ISI^{-1}$ differs across different CS-light pattern cases. The spiking pattern for dark and 480 nm illumination is shown in Fig.~\ref{CSRF2}e, similar to the biological retina (Fig.~\ref{CSRF2}b) representation, where only the real-time spiking duration (cyan shade) has been considered. The system shows no response under dark (1) or when the center and surround PCs are exposed to comparable low-intensity levels (3). However, spiking emerges (970-809 Hz) only at the same contrast but at higher intensity levels, reflecting OSHN's contrast-illumination-dependent spiking. During this, a decrease in spiking rate over time can be observed, reflecting spike frequency adaptation (SFA) to avoid persistent stimuli and also helping reduce energy consumption. This could be attributed to the implementation of the Ca-channel leaky integrator. \cite {Thakar2023} When an intensity difference exists, the OSHN remains silent if the surround is brighter (4) but spikes when the center is brighter (614 Hz for C-illuminated ($I_1$)/S-dark and 772 Hz for C-illuminated ($I_2$)/S-illuminated ($I_1$), 2), demonstrating CS-contrast-dependent spiking, facilitating background suppression and making it suitable for edge detection. A similar study has been carried out under dark and 800 nm illumination, showing a contrast-illumination-dependent $V_M$ in Supporting Information 8 (Fig. S8a). The extracted raster plot is shown in Fig. S8a, and the retina-like representation with maximum spiking rates is shown in Fig.~\ref{CSRF2}f. Further, the CSRF-mediated RGC spatiotemporal action potential, $V_M$, is shown in Supporting Information 9 for step changes in $I_{syn}$ across CS-light patterns of dark and 480 nm illumination, demonstrating sensitivity-controlled RGC spiking exhibiting VA under dynamic illumination.

High-acuity color and achromatic vision in the human retina are supported by short-, mid-, and long-wavelength cones. Midget RGCs that form the majority of RGCs, receive direct paired input from L- and M-cones, thereby exhibiting spectral-spatial opponency.\cite{Patterson2019, Kim2021} Detailed action potentials, raster plots, and extracted spiking rates are provided in Fig. S8b. Due to green (480 nm)/ red (800 nm) color antagonism (Fig.~\ref{CSRF2}g), OSHN shows a baseline firing rate of 364 Hz for C-green and 286 Hz for C-red (1); in contrast, only surrounding illumination (S-red or S-green) produces no spiking (2). However, center illumination at 800 nm surrounded by 480 nm results in an elevated spiking rate (964 Hz, 3) than in the opposite configuration (386 Hz, 4), demonstrating dominant ON-C-red/OFF-S-green behavior. This shows OSHN’s higher spiking rate than the retina and its ability to encode luminance contrast and ON-L/OFF-M chromatic opponency, thereby improving color discrimination and facilitating robust, faster perception of camouflaged visual scenes.

\begin{figure} [H]	
{\includegraphics[width=1\textwidth]{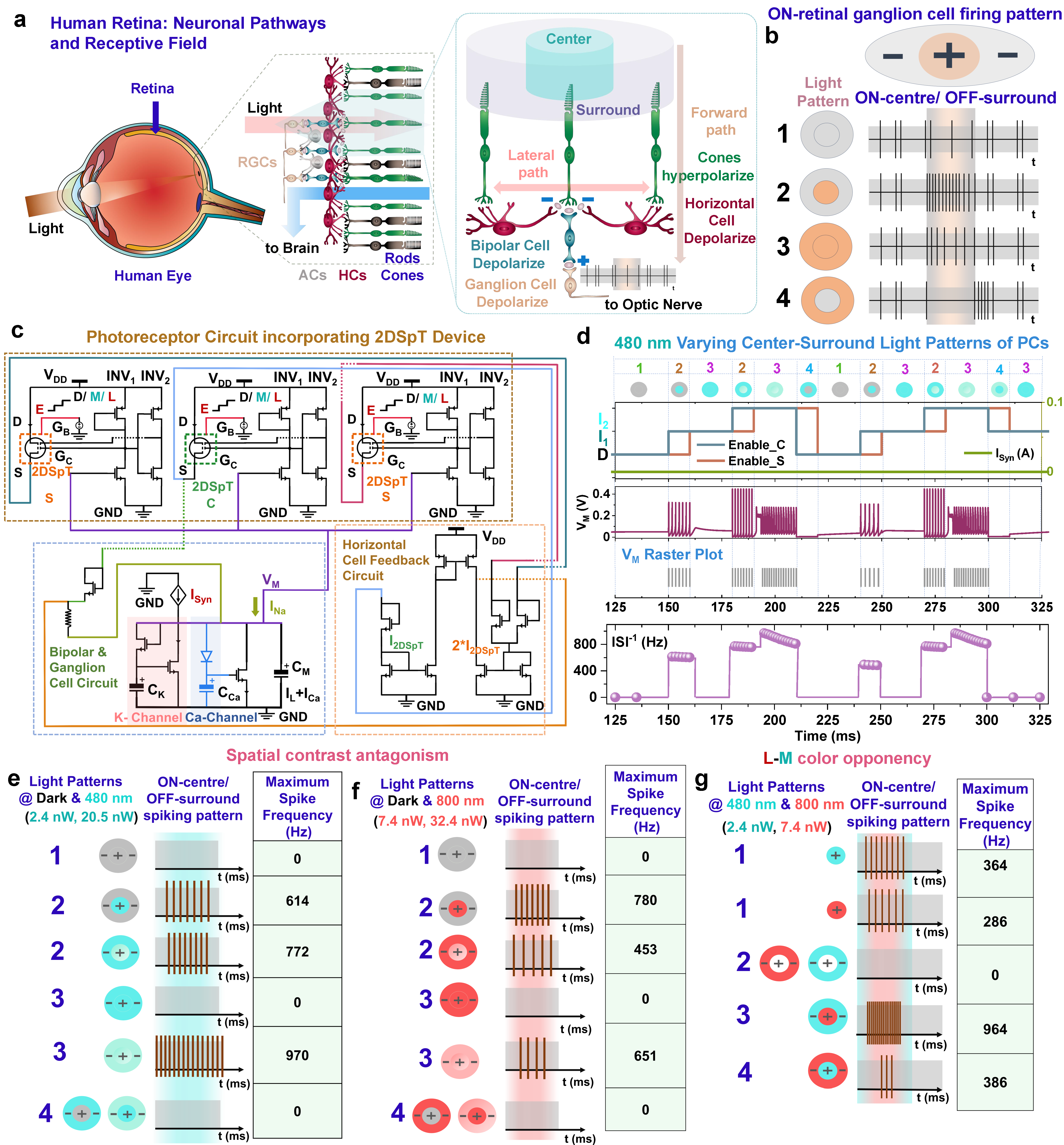}}
\caption{\textbf{Biomimetic center-surround receptive field (CSRF) mediated ON-RGC spiking.} (a) Light interpretation path of human retina (light$\rightarrow$RGC$\rightarrow$BC$\rightarrow$PC$\rightarrow$synaptic response$\rightarrow$BC$\rightarrow$RGC$\rightarrow$optic nerve) and CSRF field of RGC highlighting PCs, BCs, HCs, ACs, and ganglion cells. (b) Schematic of biological ON-center/OFF-surround RGC spiking for different dark and illumination patterns. (c) Circuit schematic used to emulate the ON-center/OFF-surround receptive field of RGCs with lateral inhibition to regulate the forward retinal response. (d) Action potential of OSHN in response to the center-surround light pattern of the PCs, corresponding extracted raster plot and spiking rates under dark and varying intensities of 480 nm illumination. (e) Biologically inspired raster plot representation of the action potential shown in (d), scaled and drawn to reflect the temporal characteristics of the spiking events. (f) Similar bio-inspired spiking representation under dark and at two different 800 nm illumination intensities. Further, (g) shows the midget RGCs' ON-L/OFF-M chromatic opponency emulation and corresponding differential bio-inspired spiking patterns at 480 nm and 800 nm.}
\label{CSRF2}
\end{figure}


\subsection{\textbf{Bio-plausible SNN with CSRF-mediated RGC Circuit for Camouflaged Object Detection}}

COD is a critical computer vision task focused on the identification and segmentation of objects (or foreground) that blends well with the background and often forms the bottleneck for traditional vision systems. Sophisticated and computationally intensive techniques such as search and identification networks (SINet) \cite{fan2020camouflaged, fan2022advances} and vision transformers \cite{yang2021uncertainty} have been proposed to efficiently handle COD workloads. Here, we demonstrate a bio-plausible SNN COD approach augmented with the retinomorphic CSRF-mediated RGC spiking circuit, OSHN, that significantly outperforms traditional SNN frameworks. The developed hardware-aware SNN simulation framework (Fig.~\ref{COD}a) consists of four bio-realistic preprocessing stages to mimic the behavior of the CSRF-mediated RGC spiking circuit: (1) a photoreceptor layer calibrated to the spiking behavior (Fig.~\ref{CSRF2}d and Fig. S8a) of OSHN at 480 nm and 800 nm (methodological details are provided in Supporting Information 10: photoreceptor layer), (2) a center-surround receptive field layer including difference-of-Gaussians (DoG) with surround weight, $w_s$ = 0.8845 (Supporting Information 10: CSRF layer), derived from spiking frequency (Fig.~\ref{CSRF2}d and Fig. S8a) of 800 nm and 480 nm, (3) a visual adaptation layer implementing scotopic and photopic adaptation (Supporting Information 10: VA layer). The VA layer adds no learnable parameters. It applies a spatial gain map: low light regions receive a sensitivity boost with scotopic adaptation time constant, $\tau_s$ = 30 ms ($ISI^{-1}$(t) rise from 306 Hz to 650 Hz, Fig.~\ref{adap}c) and bright regions receive compression with photopic adaptation time constant, $\tau_p$ = 27 ms ($ISI^{-1}$(t) falls from 603 Hz to 271 Hz, Fig.~\ref{adap}e). Finally, (4) a spike encoder with piecewise-linear probability mapping calibrated (Supporting Information 10: spike encoder) to the differential 480 nm and 800 nm spiking frequency (Fig. S8b) is used. Parameter values used in different stages, along with their calibration source and biological significance are provided in Table S4. Furthermore, the neurons were trained in the SNN framework using the ATan surrogate gradient method \cite{fan2022advances} with AdamW optimizer and cosine annealing (Supporting Information 10, SNN architecture and training) on three different datasets with progressive complexity: binary classification on the FMNIST dataset, camouflaged object detection on the COD10K-v3 \cite{fan2022advances}, and a synthetic camouflaged dataset generated via octave fractal Brownian motion noise backgrounds and alpha-composited targets in accordance with the synthetic camouflaged object detection (SCODE) paradigm. \cite{zhang2023camouflaged} Performance of this SNN augmented with the CSRF-mediated circuit has been benchmarked against a conventional SNN (without any photoreceptor layer or receptive field or adaptation layer) with the same network hyperparameters and training methodology (Fig.~\ref{COD}b). 

A comprehensive analysis indicates that the SNN with CSRF-mediated vision capability exhibits significant improvement in classification accuracies by 4.4$\%$, 10.4$\%$, and 28.4$\%$ for the FMNIST, COD10K-v3, and synthetically generated camouflaged datasets, respectively, as compared to the conventional SNN (Fig.~\ref{COD}c). The substantial improvements in classification accuracies obtained for the complex camouflaged datasets are attributed to the preprocessing layers, specifically the DoG center-surround filter, which mimics CSRF-mediated processing, exaggerates local spatial derivatives, and encodes them as a learnable, spike-accessible signal by converting suppressed boundary contrast (similar to ganglion cells). On the contrary, traditional SNN encoders convert pixel intensities into spike probabilities or Bernoulli rates, resulting in low contrast boundaries that are indistinguishable in spike space. \cite{fang2021incorporating} Therefore, the retinomorphic CSRF-mediated RGC circuit demonstrated in this work could serve as a compact (area-efficient) and energy-efficient biomimetic alternative to prevailing complex, computationally intensive techniques for COD.

\begin{figure} [H]	
{\includegraphics[width=1\textwidth]{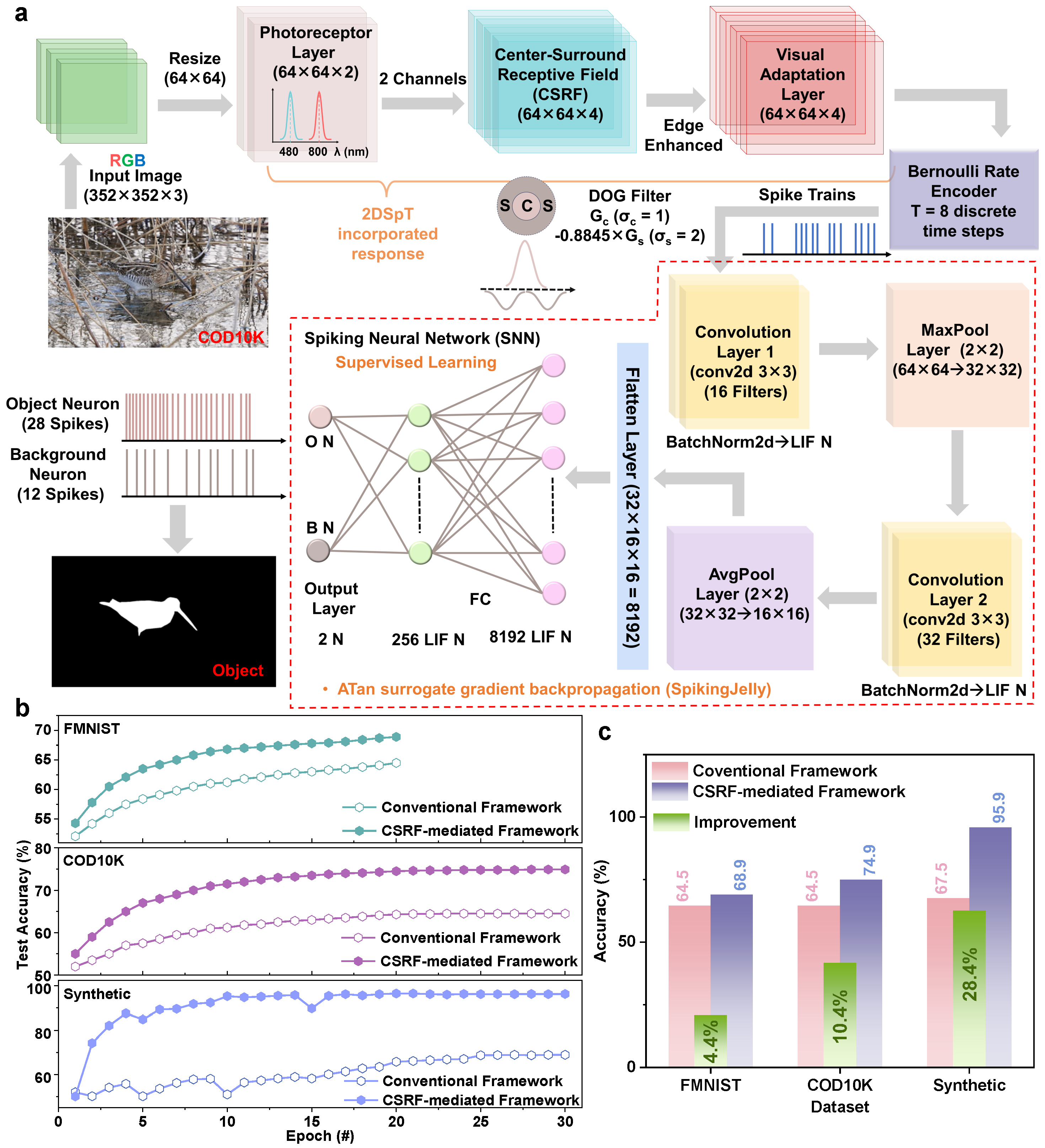}}
\caption{\textbf{Camouflaged object detection (COD) with conventional and CSRF-mediated SNN frameworks.} (a) Schematic of the CSRF-mediated SNN framework (highlighting photoreceptor, CSRF, and visual adaptation layers). (b) Test accuracy as a function of the number of epochs for conventional and CSRF-mediated SNN frameworks on the FMNIST, COD10K, and the synthetic datasets. (c) Comparison of maximum test accuracy attained by the conventional and the CSRF-mediated SNN framework on the FMNIST, COD10K, and the synthetic datasets, and the percentage improvement.}
\label{COD}
\end{figure}

\section*{Conclusion}\label{sec13}

This work presents an energy-efficient optical spiking neuron, OSHN, that emulates key preprocessing functionalities of the human retina. OSHN incorporates a 2DSpT that exhibits wavelength- (480 nm and 800 nm) and intensity-dependent bell-shaped anti-ambipolar subthreshold transfer characteristics. Leveraging its light-modulated bell-shaped behavior, OSHN demonstrates WIS spike encoding with energy consumption as low as 0.9 pJ (D), 2 pJ (480 nm), and 24.5 pJ (800 nm) at low (biological)-to-high spiking rate (0 - 2 kHz) and response times (4.2 \textmu s - 1.25 ms) substantially faster than the human retina (30 ms - 60 ms), enabling rapid decision-making capability (Table S6). Ca$^{2+}$-mediated visual adaptive capability (at 480 nm) of OSHN prevents system saturation and ensures robust perception across a wide illumination intensity range. In addition, OSHN mimics CSRF-dependent RGC spiking (under dark, 480 nm and 800 nm) and midget ganglion cells' L-M cone color opponency, where CSRF-mediated retinal preprocessing enhances contrast while accentuating edges and spatial transitions, and color opponency enables concurrent chromatic-spatial visual perception. The above findings represent important steps towards practical applications, such as COD, for evaluating system-level efficacy. A CSRF-mediated SNN framework has been developed and trained on FMNIST, COD10K, and synthetic camouflaged datasets, achieving test accuracies of 68.9$\%$, 74.9$\%$, and 95.9$\%$, respectively. The CSRF-augmented SNN exhibits substantial improvements in classification accuracy of 4.4$\%$, 10.4$\%$, and 28.4$\%$ for FMNIST, COD10K, and synthetic camouflaged datasets, respectively, compared to a conventional SNN. A comprehensive performance comparison of OSHN is provided in Table S5 (non-spiking perception systems) and Table S6 (spiking perception systems), along with the parameters for the human biological retina. Collectively, these findings highlight the potential of 2DSpT-based neuromorphic architecture for energy- and area-efficient, low-latency retinomorphic systems with high acuity and spectral discrimination.

\section*{Methods}

\subsection*{\textbf{Device Fabrication}}

The device was fabricated on a p$^+$-Si/SiO$_2$ (285 nm) substrate. The three coplanar back electrodes: one channel gate ($G_C$) and the two shorted barrier gates ($G_B$s) were patterned using electron beam lithography (EBL, Raith 150-Two) followed by metallization (Ti (4 nm)/Au (30 nm) in AJA sputter system) and lift-off. Single crystals of hBN (dielectric) and WSe$_2$ (2D channel material) were procured from SPI Supplies. The hBN and WSe$_2$ flakes were micromechanically exfoliated using adhesive scotch tape and transferred from the scotch tape to a polydimethylsiloxane (PDMS) stamp. First, the hBN flake was focus-transferred onto the prefabricated gate patterns under the Olympus BX-63 microscope, followed by WSe$_2$ flake dry-transfer. Next, the source/drain (S/D) contacts were patterned using EBL, making an overlap with the $G_B$s on both sides. Finally, contact metals (Cr (4 nm)/Pt (30 nm)/Au (70 nm)) were deposited in an AJA sputter system, and liftoff was performed to complete device fabrication.

\subsection*{\textbf{Device Characterization}}

AFM measurements for WSe$_2$ and hBN flakes were carried out using an MFP-3D system (Oxford Instruments). Raman and PL spectra were collected using a LabRAM HR Evolution-RAMAN system. Optoelectronic measurements were performed in ambient conditions using a Keysight B1500A semiconductor device analyzer. The device was wire-bonded prior to optoelectronic characterization using a TPT HB05 wire bonder. Photoresponse was characterized using an NKT SuperK EXU-6 laser coupled to a BX-63 Olympus microscope, with the optical power systematically adjusted using HOLMARC optical density filters.                                

\subsection*{\textbf{Circuit Simulations and Analysis}}

Circuit simulations were performed using Spectre (Cadence). Scaled experimental data from the 2DSpT measurements were incorporated via a lookup-table-based Verilog-A model. The transistors and capacitors were implemented using a general-purpose 45 nm process design kit (gpdk045). Simulation outputs were exported as comma-separated value (CSV) files and subsequently analyzed for parameter extraction using MATLAB and Python.

\section*{\textbf{Data availability}}

Source data is available from the corresponding author upon request.

\section*{Acknowledgements}

The authors thank Dr. Bebhash S Raj for insightful discussions. The authors express their gratitude to the Indian Institute of Technology Bombay Nano Fabrication Facility (IITBNF) and the 2D Materials and Devices Lab for providing access to facilities for device fabrication and characterization. S.S. acknowledges the support of the Prime Minister’s Research Fellowship (PMRF) Ph.D. scheme, Government of India. S.L. acknowledges funding support from the Department of Science and Technology, under the project grant DST/NM/TUE/QM - 8/2019 (G)/2 and National Quantum Mission, an initiative of the Department of Science and Technology, Government of India.

\bibliography{mainref}

\end{document}


\onehalfspacing


\title[Article Title]{\centering
\textbf{Supporting Information}\\
A Retinomorphic Optical Spiking Neuron for Camouflaged Object Detection}


\author[1]{\fnm{Srilagna} \sur{Sahoo}}\email{srilagna.sahoo@gmail.com}

\author[2]{\fnm{Adwaaiit} \sur{Pande}}\email{adwaaiit.pande@gmail.com}

\author[3]{\fnm{Kartikey} \sur{Thakar}}\email{kartik.iitb.ue@gmail.com}

\author[2]{\fnm{Shubham} \sur{Sahay}}\email{ssahay@iitk.ac.in}

\author*[1]{\fnm{Saurabh} \sur{Lodha}}\email{slodha@ee.iitb.ac.in}

\affil*[1]{\orgdiv{Electrical Engineering}, \orgname{Indian Institute of Technology Bombay}, \orgaddress{\street{Powai}, \city{Mumbai}, \postcode{400076}, \state{Maharashtra}, \country{India}}}

\affil[2]{\orgdiv{Electrical Engineering}, \orgname{Indian Institute of Technology Kanpur}, \orgaddress{ \city{Kanpur}, \postcode{208016}, \state{Uttar Pradesh}, \country{India}}}

\affil[3]{\orgname{Texas Instruments}, \orgaddress{\city{Bengaluru}, \postcode{560093}, \state{Karnataka}, \country{India}}}

\renewcommand{\thefigure}{S\arabic{figure}}
\renewcommand{\theequation}{S\arabic{equation}}
\renewcommand{\thetable}{S\arabic{table}}

\maketitle

\newpage
\section{\textbf{Supporting Information 1:}}

\begin{figure} [h]	
{\includegraphics[width=1\textwidth]{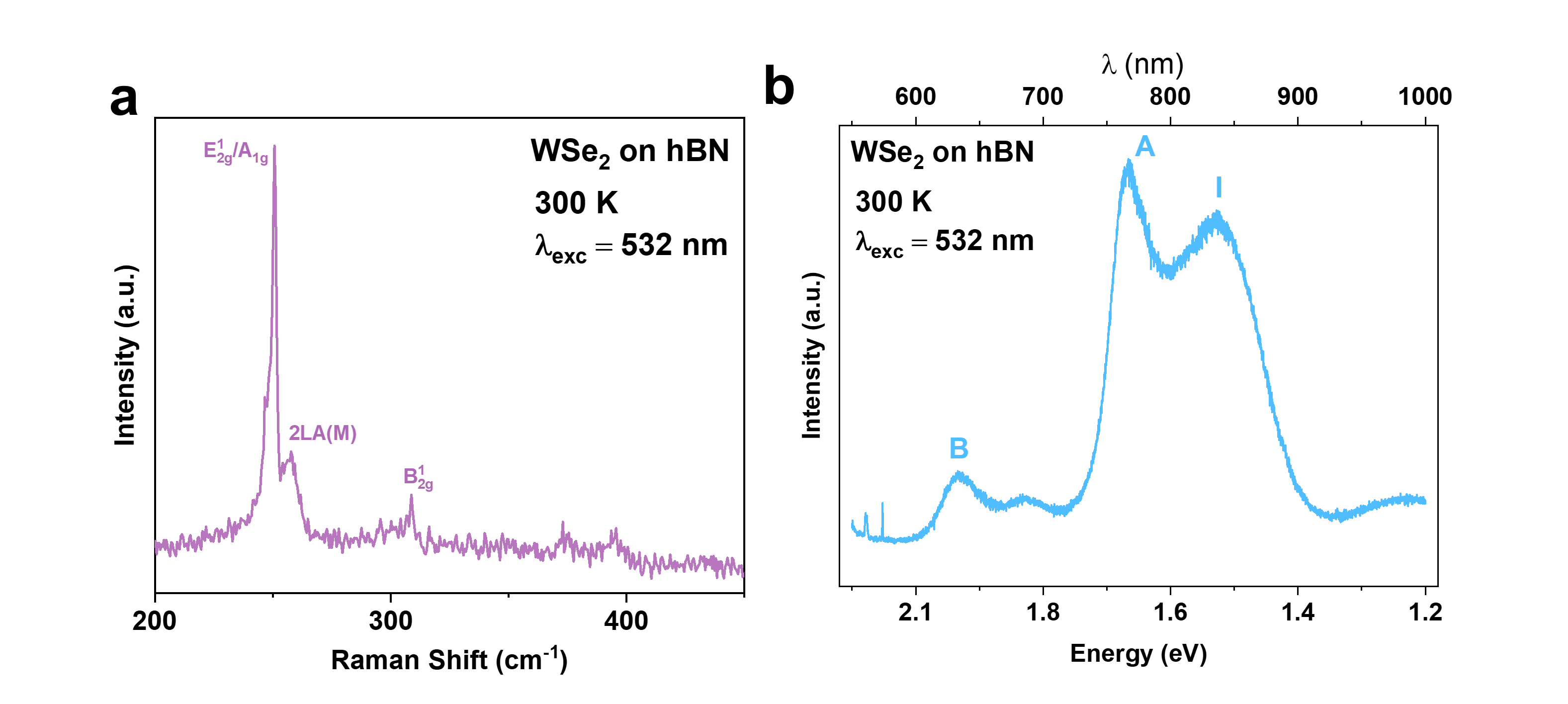}}
\caption{\textbf{Raman and photoluminescence spectra of WSe$_2$ and hBN flakes.} (a) Raman and (b) photoluminescence spectra acquired from WSe$_2$ and hBN flakes of the 2DSpT device.}
\label{SI1}
\end{figure}
\newpage
\section{\textbf{Supporting Information 2:}}

\begin{figure} [h]	
{\includegraphics[width=1\textwidth]{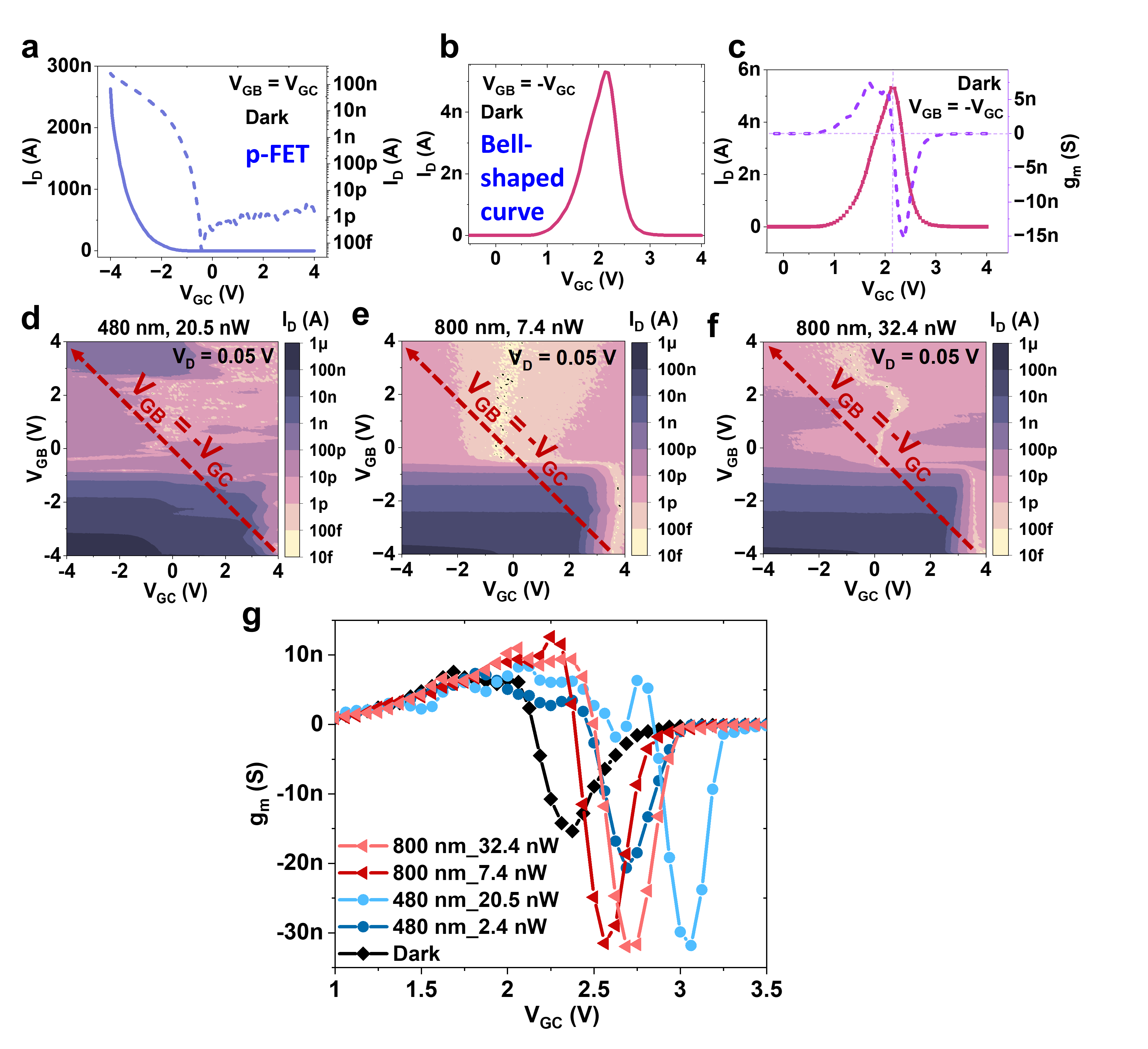}}
\caption{\textbf{Transfer characteristics (p-FET, anti-ambipolar), transconductance curves and optoelectronic drain current color maps of 2DSpT.} 2DSpT has (a) p-FET characteristics when $V_{GB}$ = $V_{GC}$ and shows (b) bell-shaped transfer characteristics with (c) positive/negative transconductance, $g_m$ ($V_{GB}$ = -$V_{GC}$) under dark. Color maps show 2DSpT drain currents as a function of back gate biases ($V_{GB}$ and $V_{GC}$) under (d) 480 nm (20.5 nW), and 800 nm ((e) 7.4 nW, and (f) 32.4 nW) illumination at drain bias, $V_D$ = 50 mV, respectively. (g) The extracted transconductance, $g_m$, under all illumination conditions.}
\label{SI2}
\end{figure}
\newpage
\section{\textbf{Supporting Information 3:}}

\begin{figure} [h]	
{\includegraphics[width=1\textwidth]{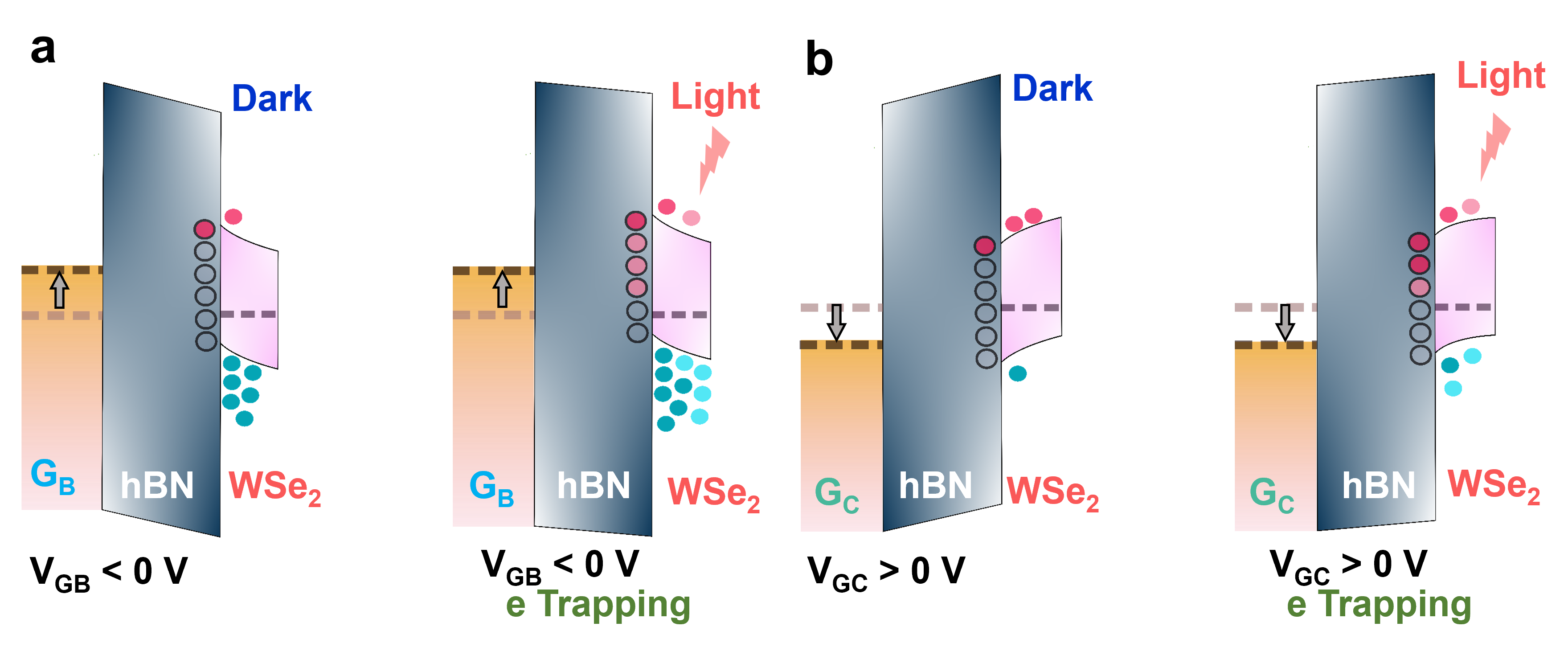}}
\caption{\textbf{Charge trapping device physics.} Charge trapping under dark and illuminated states at WSe$_2$/hBN interface due to (a) barrier gate bias, $V_{GB}$ and (b) channel gate bias, $V_{GC}$.}
\label{SI2}
\end{figure}


\newpage
\section{\textbf{Supporting Information 4:}}

\begin{figure} [h]	
{\includegraphics[width=0.9\textwidth]{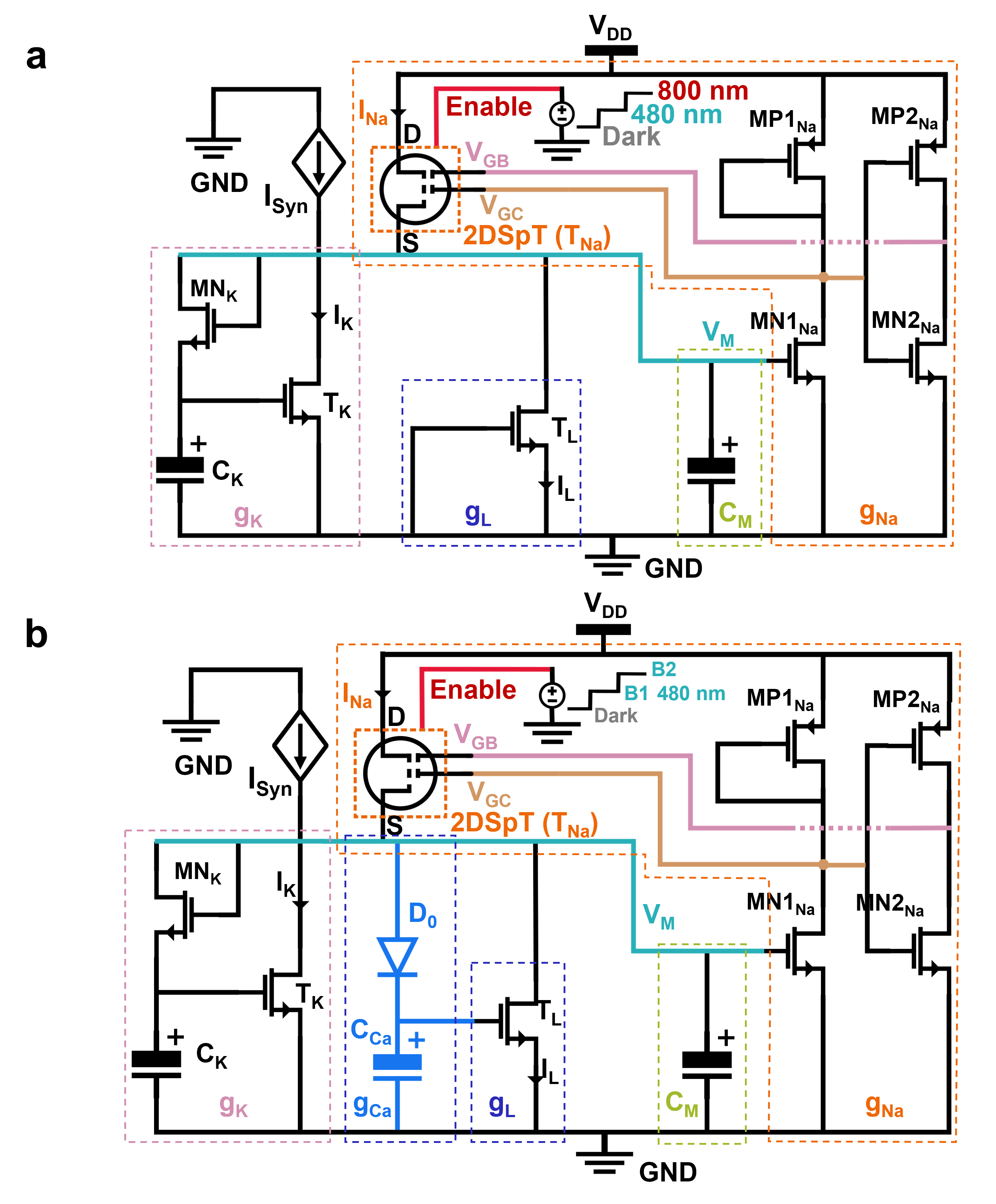}}
\caption{\textbf{Schematic of the optical spiking Hodgkin–Huxley (HH) neuron (OSHN) circuit.} The simulated optical spiking neuron circuit (a) without and (b) with a Ca ($D_0$-$C_{Ca}$) channel.}
\label{SI2}
\end{figure}

\subsection{Circuit Components}

\textbf{2DSpT:}

The phototransistor, 2DSpT, exhibits photo-modulated bell-shaped anti-ambipolar transfer characteristics that resemble the intrinsic temporal anti-ambipolar conductance of biological sodium (Na) channels. The schematics shown in Fig. S4 illustrate a simulated optical spiking Hodgkin–Huxley (HH) neuron (OSHN) circuit that integrates 2DSpT to mimic the light-gated sodium (Na) channel current.

\subsection{Integration of 2DSpT in OSHN:}

A lookup table-based Verilog-A behavioral model was implemented to capture 2DSpT's anti-ambipolar behavior. Apart from S/D and two gate biases ($V_{GC}$ and $V_{GB}$), another node Enable has been modeled to control the illumination condition that drives the 2DSpT's current during the circuit operation. To accommodate the nominal voltage range (0-1 V) of 45 nm CMOS technology, the gate biases are scaled  as follows:

\begin{equation}
{V_{GC}}' = \frac{V_{GC}}{2} - 0.85
\end{equation}

\begin{equation}
{V_{GB}}' = \frac{V_{GB}}{2} + 1.25
\end{equation}


\subsection{Other 45 nm CMOS components: }

\begin{table}[htbp]
\caption{List of components other than 2DSpT for the circuits in Supporting Information 4.}
    \centering
    \begin{tabular}{|p{2.2cm}|p{4cm}||p{3cm}|p{3cm}|p{1cm}|}\hline
         Components&  \multicolumn{3}{c|}{Dimensions/Value }& Unit\\\hline
         &  Wavelength- and Intensity- Encoding (without Ca-channel), Fig. S4a
&  \multicolumn{2}{p{5.5cm}|}{Visual Adaptation (with Ca-channel), Fig. S4b}& \\\hline
         &  &  (Scotopic)
&  (Photopic)
& \\\hline
         $V_{DD}$&  1&  1&  1& V\\\hline
         $I_{syn}$&  100-1000&  300-1800&  400-22& pA\\\hline
         $MP1_{Na}$&  120/45 (W/L)&  120/45 (W/L)&  120/45 (W/L)& nm\\\hline
         $MP2_{Na}$&  120/45 (W/L)&  120/45 (W/L)&  240/45 (W/L)& nm\\\hline
         $MN1_{Na}$&  360/45 (W/L)&  480/45 (W/L)&  1320/45 (W/L)& nm\\\hline
         $MN2_{Na}$&  120/45 (W/L)&  120/45 (W/L)&  120/45 (W/L)& nm\\\hline
         $T_L$&  600/45 (W/L)&  720/45 (W/L)&  240/45 (W/L)& nm\\\hline
 $T_K$& 600/45 (W/L)& 1080/45 (W/L)& 600/45 (W/L)&nm\\\hline
 $MN_K$& 360/45 (W/L)& 120/45 (W/L)& 120/45 (W/L)&nm\\\hline
 $C_M$& 50
& 50& 50
&fF\\\hline
 $C_K$& 200& 200& 200
&fF\\\hline
 $D_0$& -& 200/200 (W/L)& 200/200 (W/L)
&nm\\\hline
 $C_{Ca}$& -& 400& 400&fF\\ \hline
    \end{tabular}
    
    \label{tab:placeholder}
\end{table}

\subsection{Circuit operation:}

The source (S) node integrates the synapting current, $I_{Syn}$, and charges the membrane capacitance, $C_M$, while leaking through $T_L$. As $C_M$ continues to charge and $V_M$ continues to potentiate until it is sufficient to activate $MN1_{Na}$, it pulls down $V_{GC}$ and pulls up $V_{GB}$, and 2DSpT is turned on. This results in a large sodium ion current, $I_{Na}$, which shoots up $V_M$ and also activates the potassium ion current, $I_{K}$, with a delay. The activation of $T_{K}$ causes the discharge of $C_M$ and a decrease in $V_M$, causing a spike (action potential) (Fig. S4a). In Fig. S4b, the additional Ca ($D_0$-$C_{Ca}$) channel integrates the leakage current, leading to a decrease in the spike rate over time, indicating adaptive behavior. \cite{Thakar2023, Beck2020}

\newpage
\section{\textbf{Supporting Information 5:}}

\begin{figure} [htbp]	
{\includegraphics[width=1\textwidth]{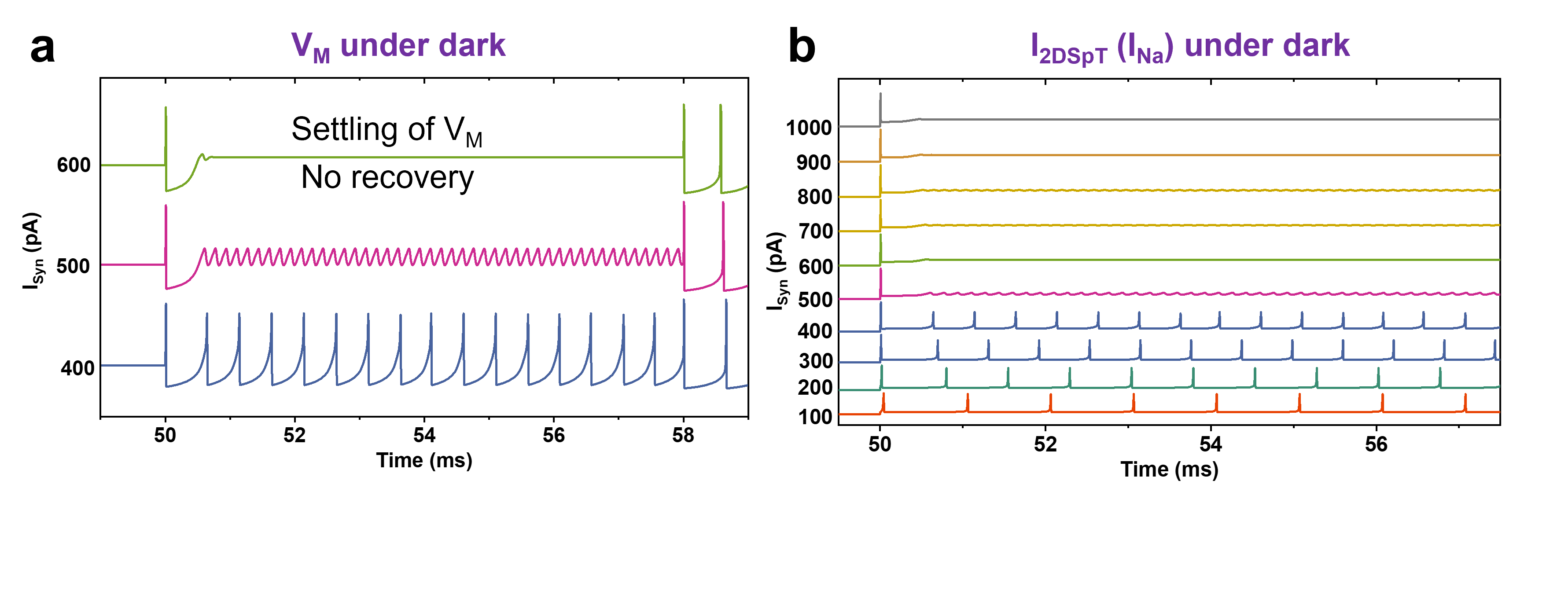}}
\caption{\textbf{Action potential and Na-channel current under dark.} (a) Action potential, $V_M$ at $I_{Syn}$ = 400 pA, 500 pA, and 600 pA, and (b) $I_{Na}$ at $I_{Syn}$ = 100 pA - 600 pA under dark.}
\label{SI2}
\end{figure}

\newpage
\section{\textbf{Supporting Information 6:}}

\begin{figure} [htbp]	
\centering
{\includegraphics[width=1\textwidth]{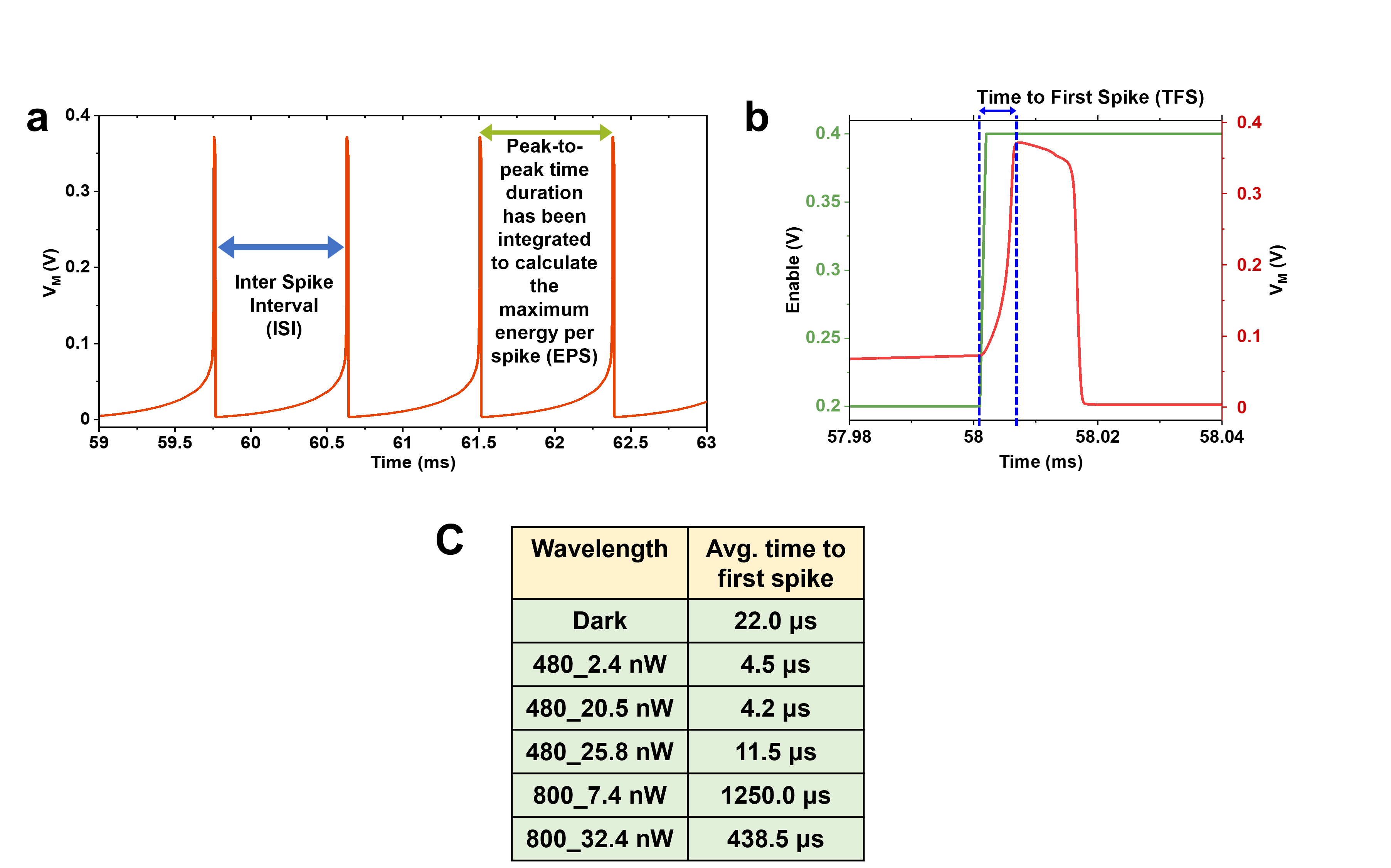}}
\caption{\textbf{Calculation of spike properties.} (a) Action potential, $V_M$ at $I_{Syn}$ = 100 pA under 480 nm (2.4 nW) illumination, highlighting the inter spike interval ($ISI$) and the integrated time duration for the evaluation of maximum energy per spike (EPS). (b) Elapsed time from the stimulus onset to the first spike (TFS). (c) Average latency to the first spike after the stimulus onset was extracted from Fig. 3a for different illumination.}
\label{SI2}
\end{figure}


\subsubsection{Maximum energy per spike (EPS):}


\begin{equation}
\mathrm{EPS} = \int \left( V_{DD}\, I_{\mathrm{net}} + V_M\, I_{\mathrm{Syn}} \right)\, dt
\end{equation}

where $I_{net}$ is the net current drawn by the entire circuit at node $V_{DD}$.

\newpage
\section{\textbf{Supporting Information 7:}}

\textbf{CSRF arrangement of human retina:}

The massive convergence from photoreceptor cells (PCs) (80–110 million rods and 4–5 million cones, with a peak distribution at the fovea) to bipolar cells (BCs, 36 million) and the ultimate integration onto fewer retinal ganglion cells (RGCs) (1–2 million/eye) exemplify the retina’s highly parallel processing architecture. \cite{Feher2012} Parallel processing in the retina involves simultaneous segregation and encoding of multiple visual features into distinct neural paths that are analyzed independently. It initiates at the photoreceptors (rods and cones), the first-order neurons, where black-and-white contrast and color are separated according to their wavelength sensitivities. Further, the PC signals are transmitted to two types of BCs (ON and OFF), second-order neurons, depending on the contrast between the foreground (bright or dark) and the background luminance (which will be discussed in more detail later).\cite{Ichinose2022} Signals from ON and OFF BCs are subsequently transmitted to corresponding ON and OFF RGCs (third-order neurons), while some RGCs integrate both ON–OFF type BCs. These ganglion cell pathways then project to distinct brain targets for visual perception via the optic nerve.\cite{Ichinose2022} The RGCs utilize a CSRF arrangement shaped by lateral inhibition from horizontal and amacrine cells. The receptive field has a central disk and a concentric surrounding ring, allowing RGCs to highly prioritize spatial contrast, which specifies the edges of objects.\cite{Turner2018Receptive, Gaynes2022}

\begin{figure} [h]
\centering
\includegraphics[width=0.7\textwidth]{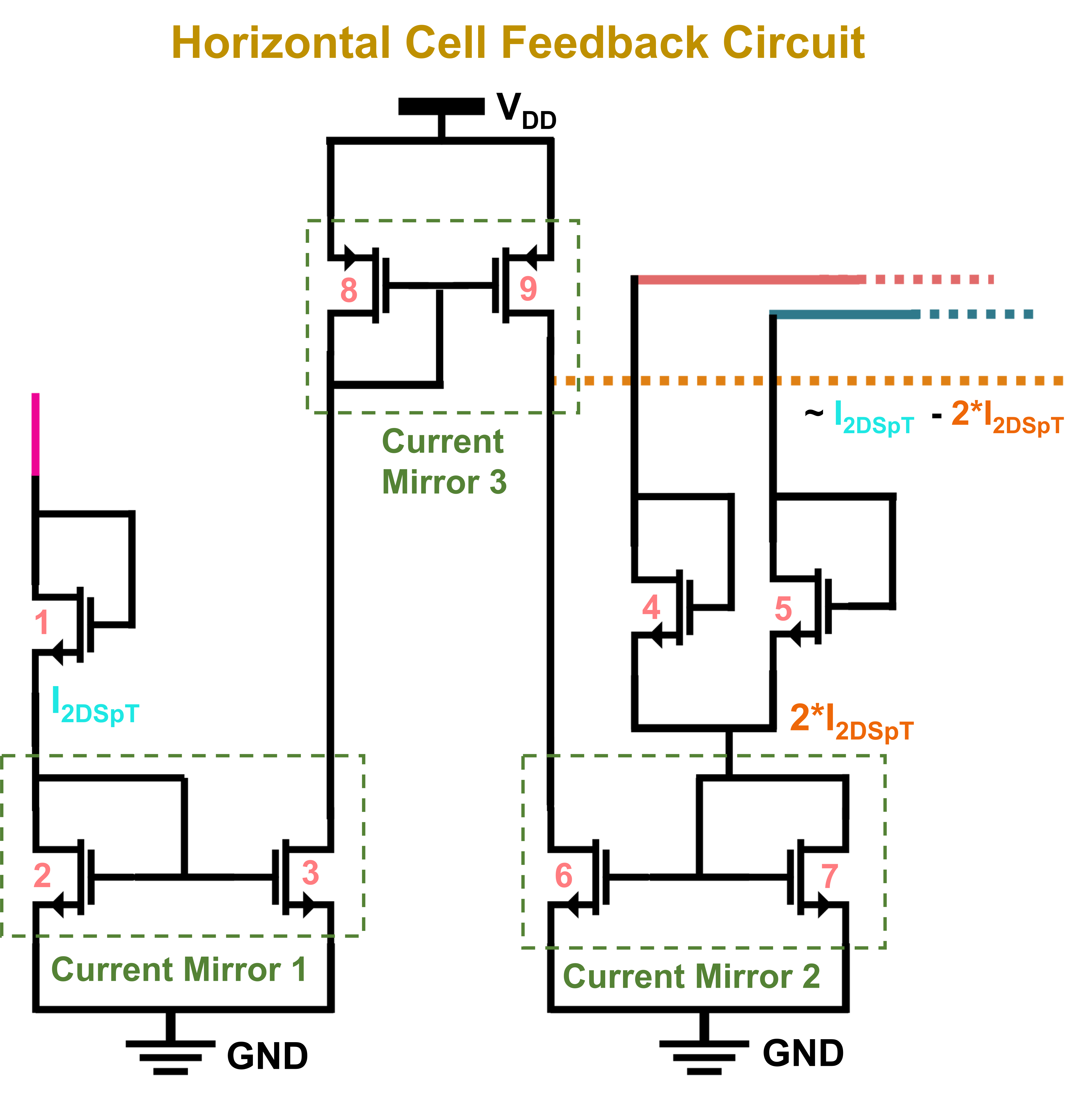}
\caption{\textbf{Biomimetic horizontal cell circuit.} The current steering circuit (mimicking a horizontal cell) is optimized in the saturation region. It consists of 3 current mirrors (CMs). CM1 mirrors the center-2DSpT current, and CM2 mirrors the summed currents of two surrounding-2DSpTs onto CM3. Finally, CM3 feeds the subtracted current output to the retinal ganglion cell (RGC) spiking circuit through a resistor (mimicking a bipolar cell).}
\label{SI2}
\end{figure}
\newpage
\noindent \textbf{Table S2:} List of components for the current steering circuit (horizontal cell feedback circuit) in Fig. 5c.
\begin{center}
    \centering
    \begin{tabular}{|c|c|c|}\hline
         Components & Dimension/Value &Unit \\\hline
         1&  120/45 (W/L)& nm\\\hline
         2&  480/45 (W/L)& nm\\\hline
         3&  480/45 (W/L)& nm\\\hline
         4&  120/45 (W/L)& nm\\\hline
         5&  120/45 (W/L)& nm\\\hline
         6&  960/45 (W/L)& nm\\\hline
         7&  960/45 (W/L)& nm\\\hline
         8&  240/45 (W/L)& nm\\\hline
         9&  240/45 (W/L)& nm\\\hline
    \end{tabular}
    
    \label{tab:placeholder}
\end{center}
\noindent \textbf{Table S3:} List of components other than 2DSpT for the circuit in Fig. 5c.
\begin{center}
    \centering
    \begin{tabular}{|p{2.2cm}|p{4cm}||p{3cm}|p{1cm}|}\hline
         Components&  \multicolumn{2}{|c|}{Dimensions/Value }& Unit\\\hline
         &  480 nm and 800 nm individually (Fig. 5e-f)&  480 nm and 800 nm color opponency (Fig. 5g)& \\\hline
         $V_{DD}$&  1&    1& V\\\hline
         $I_{syn}$&  0&    0& pA\\\hline
         $MP1_{Na}$&  120/45 (W/L)&   120/45 (W/L)& nm\\\hline
         $MP2_{Na}$&  120/45 (W/L)&    120/45 (W/L)& nm\\\hline
         $MN1_{Na}$&  120/45 (W/L)&    120/45 (W/L)& nm\\\hline
         $MN2_{Na}$&  120/45 (W/L)&    120/45 (W/L)& nm\\\hline
         $T_L$&  120/45 (W/L)&    120/45 (W/L)& nm\\\hline
 $T_K$& 120/45 (W/L)&  480/45 (W/L)&nm\\\hline
 $MN_K$& 240/45 (W/L)&  120/45 (W/L)&nm\\\hline
 $C_M$& 50
&  50
&fF\\\hline
 $C_K$& 200&  200
&fF\\\hline
 $D_0$& 200/200 (W/L)&  200/200 (W/L)
&nm\\\hline
 $C_{Ca}$& 200/200 (W/L)&  400&fF\\ \hline
    \end{tabular}
    \label{tab:placeholder}
\end{center}


\newpage
\section{\textbf{Supporting Information 8:}}

\begin{center}
{\includegraphics[width=1\textwidth,trim={0.3cm 2cm 2cm 0.5cm}, clip]{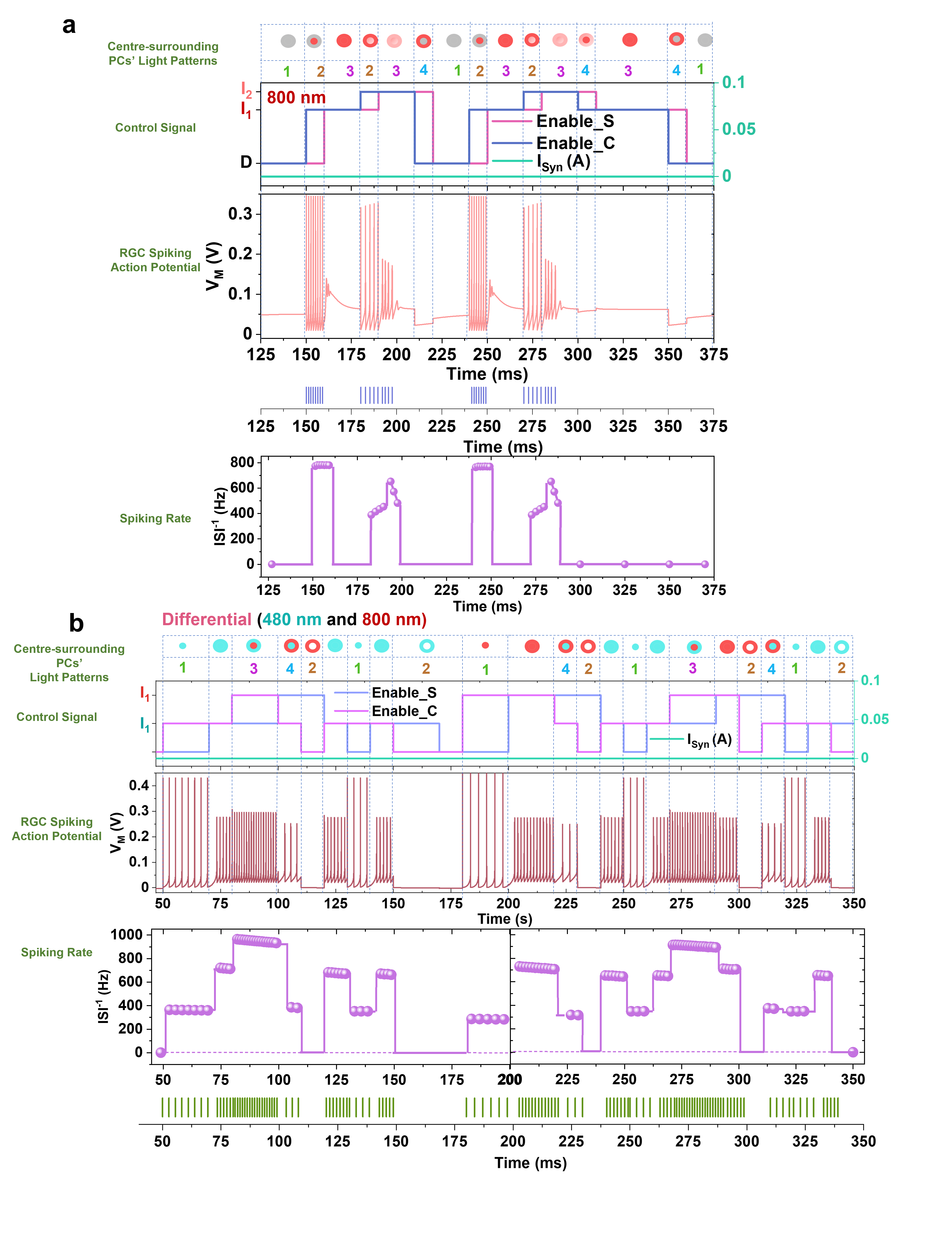}}
\end{center}

\noindent \textbf{Fig.~S8}: \textbf{Spiking pattern of OSHN with center-surround receptive field (CSRF) under 800 nm illumination and the L-M color opponency.} Biomimicking of the CSRF-mediated ON-RGC action potentials, raster plots, and spiking rates under (a) dark and 800 nm (2 intensities) illumination, and (b) differential center-surround illumination of 480 nm and 800 nm at the lower intensity of each.

\newpage
\section{\textbf{Supporting Information 9:}}


\begin{center}
\includegraphics[width=0.9\textwidth, trim=0.3cm 3cm 1cm 0.5cm, clip]{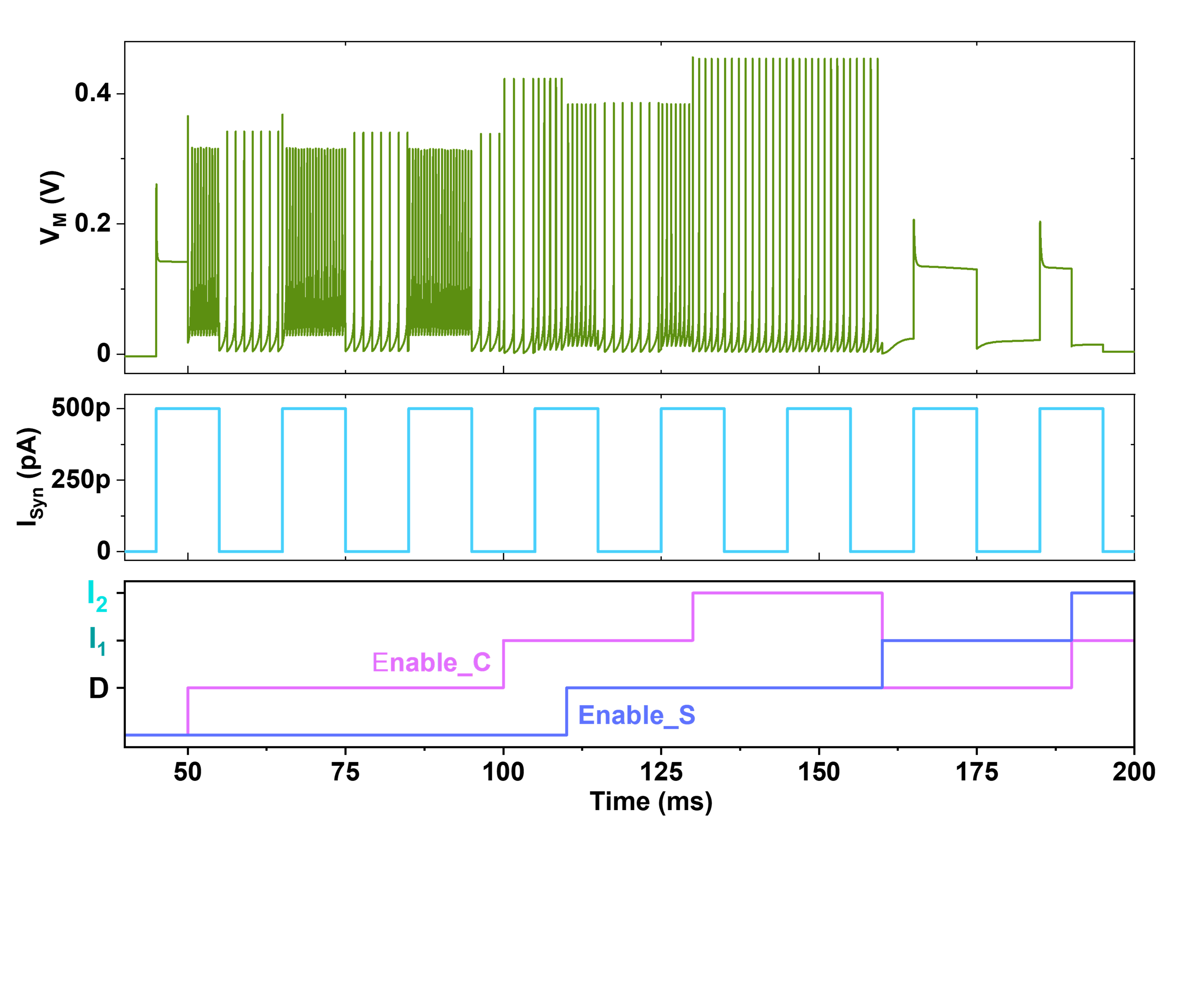}
\end{center}

\noindent \textbf{Fig.~S9}: \textbf{CSRF-mediated visual adaptation (VA) of OSHN.} Biomimicking of the CSRF-mediated VA of ON-RGC spiking circuit.

\newpage
\section{\textbf{Supporting Information 10:}}

\textbf{Device-calibrated preprocessing for camouflaged object detection.}

\subsection{SNN framework incorporating the CSRF-mediated RGC Circuit:}
The developed simulation framework for analyzing system-level performance of the CSRF-mediated RGC circuit comprises four device-calibrated preprocessing stages followed by a convolutional LIF-SNN classifier. The details regarding the preprocessing stages are as follows: 

\subsubsection{\textbf{Stage 1: Photoreceptor (PR) Layer}}

The photoreceptor layer emulates OSHN wavelength-selective log-compression at 480 nm and 800 nm. RGB is first converted to luminance (0.299R + 0.587G + 0.114B), and then the two channels corresponding to the two wavelengths are computed as follows:
\begin{equation}
ch_{i}(x, y) = log(1 + k_{i} \cdot I \cdot w_{i}) / log(1 + k_{i}), \quad i \in \{480,800\}
\end{equation}

Where the parameters are defined and obtained as follows: I(x,y) is the normalized pixel luminance at position (x,y), obtained by converting the RGB input image to a scalar grayscale value using the standard luminance formula (0.299R + 0.587G + 0.114B), with values in the range [0,1]. The index $i$ refers to the two wavelength channels: $i$ = 480 for 480 nm and $i$ = 800 for 800 nm illumination. The spectral sensitivity weights, $w_i$, are derived from the spiking frequencies at 480 nm (Fig. 5d) and 800 nm (Fig. S8a) by taking the arithmetic mean of the peaks. $w_{480}$ = 1.000 is set as the reference channel. $w_{800}$ = 0.8845 is obtained by dividing the mean spiking frequency at 800 nm (749.0 Hz) by the mean spiking frequency at 480 nm (846.8 Hz), yielding 749.0/846.8 = 0.8845. The log-compression constants $k_{480}$ = 29.203 and $k_{800}$ = 34.524 are obtained by fitting the equation $ch_{i}$(I) to the mean values of $ISI^{-1}$ at 480 nm and 800 nm (Fig. 5d, S8a) using non-linear least squares (NLS) regression with four intensity levels: dark (I = 0), dim (I = 0.2), mid (I = 0.5), and bright (I = 1.0). These constants control the degree of compressive non-linearity. A higher k value produces a steeper, more compressive log-compression curve, reflecting the stronger gain control observed in the 800 nm channel. The output of this layer is normalized to [0,1] by dividing by the fitted maximum firing rate $F_{max}$ (820 Hz for 480 nm, 735 Hz for 800 nm).

\subsubsection{\textbf{Stage 2: CSRF Layer}}

The three-photoreceptor cell CSRF ganglion circuit (OSHN, comprising left, center, and right 2DSpT, as shown in Fig. 5) characterizes the response of a single RGC at the circuit level. In the simulation, the unit computation is applied in parallel at every spatial position in the image, effectively tiling the entire image with an array of identical ganglion unit cells, each centered at a pixel, consistent with the parallel organization of retinal ganglion cells across the retinal surface. The CSRF layer is necessary because a boundary between a camouflaged object and its background is a spatial comparison between neighboring regions; it cannot be detected by a single photoreceptor operating in isolation, as in layer 1. The CSRF layer applies a difference-of-Gaussians (DoG) filter to each photoreceptor channel, implementing the ON-center response of the CSRF ganglion circuit (Fig. 5), where the ganglion fires when the center cell is brighter than the surround.

\begin{equation}
\begin{aligned}
DoG = w_c \cdot G(\sigma_{c}=1.0) - w_s \cdot G(\sigma_{s}=2.0)\\  
ON: ReLU(DoG)\\     
\end{aligned}
\end{equation}

Where $w_c$ = 1.000 is the center weight, representing the direct excitatory drive from the center 2DSpT photoreceptor to the ganglion cell, it is set to unity as the reference gain of the forward path. $w_s$ = 0.8845 is the surround weight, representing the strength of the lateral inhibitory feedback from the two surrounding 2DSpT photoreceptors through the horizontal cell circuit. $w_s$ was obtained by using the same procedure as for calibrating the photoreceptor layer parameters. The mean spiking frequency at the bright illumination level is 846.8 Hz at 480 nm and 749.0 Hz at 800 nm, which were used directly as the saturation spiking rates for the center and surround channels, respectively. The ratio 749.0/846.8 = 0.8845 gives $w_s$ as the fraction of the center response that the surround contributes at maximum illumination. G($\sigma$) denotes a 2D Gaussian kernel of width $\sigma$, normalized to unit sum. The ON-center response across both wavelength channels gives 2 output channels in total. Output: [B, 2, H, W], where B is the batch size (number of images processed simultaneously), 2 is the number of output channels (ON-center: ON-480 nm, ON-800 nm), H is the image height in pixels, and W is the image width in pixels.

\subsubsection{\textbf{Stage 3: Visual Adaptation (VA) Layer}}

For static images, temporal adaptation maps to spatial gain based on the local neighborhood mean $I_{local}$:

\begin{equation}
\begin{aligned}
G_s = 1 + 2.0\cdot(1 - exp(- I_{local}/\tau_s)), \\ G_p = exp(- 1.5\cdot max(I_{local} - 0.7,0)/\tau_p)\\
Output = Input \cdot (G_s \cdot G_p), \text{clamped to (0.1, 5.0)}
\end{aligned}
\end{equation}

Here, $G_s$ and $G_p$ are the scotopic gain functions applied to pixels in dark and bright regions of the image, modelling the sensitivity increase that occurs when the visual system adapts from bright to dim light (scotopic adaptation) and dark to bright light (photopic adaptation), respectively. $I_{local}$ is the local neighborhood mean intensity computed by averaging over a 5×5 pixel window, which serves as a proxy for the local ambient luminance level at each position. $\tau_s$ = 30 ms is the scotopic adaptation time constant, representing the rate at which the OSHN's spiking rate recovers and increases (from 306 Hz to 650 Hz in $ISI^{-1}$ vs. time curve, Fig. 4c) when illumination transitions from bright to low light condition. $\tau_p$ = 27 ms is the photopic adaptation time constant, representing the rate at which OSHN's spiking rate decreases and reaches (from 603 Hz to 271 Hz in $ISI^{-1}$ vs. time curve, Fig. 4e) a new steady state when illumination transitions from dark to bright light condition. The combined gain is clamped between 0.1 and 5.0 to remain within physiologically observed sensitivity ranges. Output: [B, 2, H, W].

\subsubsection{\textbf{Stage 4: Device-calibrated Spike Encoder}}

Finally, the spike encoder converts the adapted feature maps to binary spike trains (T = 8 time steps) via Bernoulli rate encoding. Spike probability is mapped via a piecewise-linear function calibrated to the three $ISI^{-1}$ regimes (dim, mid, and peak contrast regimes explained later) of Fig. S8b. The data contains 109 measured $ISI^{-1}$ data points recorded from the physical three-cell CSRF ganglion circuit when both 480 nm (center) and 800 nm (surround) 2DSpTs are simultaneously active, referring to the direct differential output of the ganglion circuit rather than the individual photoreceptor responses. Clustering the non-zero measured values identifies three distinct operating regimes of the circuit: a dim-contrast regime with a mean $ISI^{-1}$ of 342.7 Hz, corresponding to normalized spike probability p = 342.7$/$926.1 = 0.370), a mid-contrast regime at 686.2 Hz (p = 0.741), and a peak-contrast regime at 926.1 Hz (p = 1.000). The piecewise-linear mapping uses these values as breakpoints across three intensity segments: [0.0, 0.2] maps to p $\in$ [0, 0.370] (scotopic segment), [0.2, 0.7] maps to p $\in$ [0.370, 0.741] (transition segment), and [0.7, 1.0] maps to p $\in$ [0.741, 1.000] (photopic segment). At each of the T = 8 time steps, an independent binary spike $s_t$ is drawn as $s_t$ $~\sim$ Bernoulli (p). For the baseline model, a uniform encoder with p = I (no piecewise calibration) is used instead. The values of the fitted parameters for these layers, along with their biological significance and the source of the fit, are listed in Table S4.

\newpage
\noindent \textbf{Table S4:} Parameter values used in different stages, along with their calibration source and biological significance.
\begin{center}
    \centering
    \begin{tabular}{|p{2cm}|p{1.5cm}|p{2cm}|p{4cm}|p{4cm}|}\hline
         Parameter&  Stage&  Device value&  Calibration Source& Biological Significance\\\hline
         $w_{480}$ &  PR &  1.000 &  Fig. 5d (peak 846.8 Hz)& 480 nm sensitive PC\\\hline
         $w_{800}$ &  PR&  0.8845&  Fig. S8a (749.0/846.8 Hz)& 800 nm sensitive PC
\\\hline
         $k_{480}$ &  PR&  29.203&  Fig. 5d (NLS curve fit)& Weber-Fechner at 480 nm
\\\hline
         $k_{800}$ &  PR&  34.524&  Fig. S8a (NLS curve fit)& Weber-Fechner at 800 nm
\\\hline
         $\sigma_{c}$, $\sigma_{s}$ &  CSRF&  1.0, 2.0 px &  Device aperture geometry& Center/surround spatial extent
\\\hline
         $w_{s}$ &  CSRF&  0.8845&  Fig. 5d, S8a (800/480 nm $ISI^{-1}$ ratio& CSRF antagonism strength
\\\hline
         $\tau_s$ &  VA&  30 ms& Fig. 4c ($ISI^{-1}$ rise 300$\to$650 Hz)& Scotopic adaptation gain
\\\hline
         $\tau_p$ &  VA&  27 ms&  Fig. 4e ($ISI^{-1}$ fall 600$\to$250 Hz)& Photopic adaptation gain
\\\hline
 Encoder dim & Encoder& 342.7 Hz, p = 0.370& Fig. S8b (low-contrast $ISI^{-1}$ cluster)&Low-contrast spike threshold
\\\hline
 Encoder mid & Encoder& 686.2 Hz, p = 0.741& Fig. S8b (mid-contrast $ISI^{-1}$ cluster)&Mid-contrast spike rate
\\\hline
 Encoder peak& Encoder& 926.1 Hz, p = 1.000& Fig. S8b (peak $ISI^{-1}$)&Peak-contrast spike rate\\\hline
 T (time steps)& All& 8 & Architecture choice&Spike train temporal resolution
\\ \hline
    \end{tabular}
    \label{tab:placeholder}
\end{center}

\subsection{SNN Architecture and Training}
We utilized the leaky-integrate-and-fire (LIF) neurons which follow the simple dynamics: T$\cdot$d$V_m$/dt = -$V_m$ + R $\cdot$ $I_{in}$ and fire if $V_m$ $\geq$ $V_{th}$ and then get reset to the resting potential of the membrane The ATan surrogate gradient method (SpikingJelly) \cite{Fang2021} replaces the true derivative during the backward pass with a smooth approximation: $\frac{\partial s}{\partial V}$ $\approx$ 1/(1 + |$\alpha$ $\cdot$ $\pi$ $\cdot$ V|$^2$), where $\alpha$ controls the sharpness of the surrogate around the threshold. This approximation is differentiable everywhere and allows gradients to propagate through the spiking layers during training, while the forward pass still uses the true hard threshold. The membrane state of every LIF neuron is explicitly reset between samples using functional.reset$\_$net(). All models were trained identically using the AdamW optimizer, a variant of Adam that decouples weight decay from the gradient update for more stable regularisation, with a learning rate of 2×$10^{-3}$ and weight decay of $10^{-4}$. The learning rate was annealed following a cosine schedule over 30 epochs, smoothly decaying to zero. Cross-entropy loss was used for binary classification. Gradient norms were clipped to a maximum of 1.0 before each update step to prevent gradient explosion. A dropout rate of 0.4 was applied after the penultimate fully connected layer. For the COD10K and Synthetic datasets, a weighted random sampler corrected for class imbalance by drawing samples with probabilities inversely proportional to class frequencies. All experiments used a random seed of 42 for reproducibility.

We evaluated the performance of the developed SNN framework across three datasets. The details of the datasets are as follows: 
Fashion-MNIST is a dataset of 70,000 grayscale images of clothing items at 28×28 pixels, spanning ten categories. It was reformulated here as a binary task by grouping categories 0–4 into one class and categories 5–9 into the other. This dataset contains no camouflage and serves as a controlled baseline: accuracy improvements here confirm that the PR+CSRF benefit is not specific to camouflage but generalizes to binary classification. The 60,000 training images were divided into 50,000 for training and 10,000 for validation; the standard 10,000 test images were used for evaluation.
COD10K-v3 \cite{Fan2023} contains over 10,000 photographs of naturally camouflaged animals in their native habitats, each paired with a pixel-level binary mask. Images span 68 taxonomic categories and were originally captured at 352×352 pixels. For this study, image-level binary classification was used: images from the camouflaged animal folders were assigned label 1, while images from the non-camouflaged folder containing clearly visible objects were assigned label 0. All images were resized to 64×64 pixels to accommodate SNN memory constraints. Because the number of camouflaged and non-camouflaged images is unequal, a weighted random sampler was applied during training to ensure equal class representation per epoch. This is the primary benchmark of this study, representing real, ecologically valid camouflage under natural conditions.
The synthetic camouflage dataset \cite{zhang2023camouflaged} was generated procedurally to allow precise control over camouflage difficulty. Backgrounds were created using octave fractal Brownian motion noise \cite{Perlin1985}, a multi-scale procedural texture model that produces naturalistic patterns resembling grass, bark, and sand. Camouflaged targets were produced by blending a target texture into the background using a soft-edged shape mask M(x,y) according to:

\begin{equation}
I(x,y) = B \cdot (1 - \alpha \cdot M) + T \cdot (\alpha \cdot M) + \in, \alpha~N(0, 0.015)
\end{equation}

where B is the background texture, T is the target texture drawn from the same procedural model, M is a Gaussian-blurred binary shape mask (1 at the object location, 0 elsewhere), $\alpha$ is the blend coefficient controlling camouflage difficulty, and $\in$ is additive Gaussian sensor noise with standard deviation 0.015. When $\alpha$ is small, the target is nearly indistinguishable from the background; when $\alpha$ is large, the object is clearly visible. Three difficulty levels were defined: hard camouflage with $\alpha$ between 0.10 and 0.22 (40$\%$ of samples), in which the target is nearly invisible; medium camouflage with $\alpha$ between 0.23 and 0.40 (40$\%$ of samples), in which the target is substantially blended; and easy camouflage with $\alpha$ between 0.41 and 0.60 (20$\%$ of samples), in which the target is moderately visible. The dataset comprised 5,000 training images, 1,000 validation images, and 2,000 test images, all at 48×48 pixels.

\newpage
\newcolumntype{Y}{>{\raggedright\arraybackslash}X}

\noindent \textbf{Table S5:} Biomimicking of retinal preprocessing functionalities facilitated by various non-spiking photoactive systems.

\begin{center}
\label{tbl1}
\centering
\small
\renewcommand{\arraystretch}{1.4}
\setlength{\tabcolsep}{4pt}

\begin{tabularx}{\linewidth}{|Y|Y|Y||Y|Y|Y|Y|Y|Y||Y|}
\hline


\textbf{Ref.} &\textbf{Photo\-active material}&\textbf{Neuro\-morphic Functionality}&\multicolumn{6}{c||}{\textbf{Retinal Functionality}} &\textbf{Appli\-cation}\\
\hline

& &  & WIS & SFA &\textbf{VA}  &  \multicolumn{2}{c|}{\textbf{CSRF}}&\textbf{CSRF\-+VA}& \\
\hline
\hline
& & &   &&&Luminance contrast&Color Opponency&&\\
\hline
\hline

Adv. Fun. Mat. (2023) \cite{Gao2023} & InP QDs/ InSnZO & Optical Synapse & No & No & \cellcolor[HTML]{B3F0DF}yes & No & No &No& ANN on MNIST \\
\hline

Adv. Optical Mat. (2024) \cite{Li2024} & Organic semiconductor & Photo\-active organic FeFET &  No & No & \cellcolor[HTML]{B3F0DF}Yes & No & No &No& ANN on MNIST\\
\hline
Sci. Adv. (2020) \cite{Wang2020} & WSe$_2$ & PC and BC &  No & No & No & \cellcolor[HTML]{B3F0DF}BC RF & No &No& CNN on checkered pattern \\
\hline

ACS Nano (2022) \cite{Zhang2022} & WSe$_2$ & PC and BC &  No & No & No & No & No &No& No \\
\hline

ACS Nano (2024) \cite{Han2024} & WSe$_2$ & PC and BC  & No & No & No & \cellcolor[HTML]{B3F0DF}BC RF & No &No& Edge enhancement by CNN \\
\hline



\hline
\end{tabularx}
\end{center}

\newpage
\noindent \textbf{Table S6:} Biomimicking of multiple retinal preprocessing functionalities facilitated by various spiking photoactive architectures.
\begin{center}
\label{tbl2}
\centering
\tiny
\renewcommand{\arraystretch}{1}
\setlength{\tabcolsep}{2.2pt}

\makebox[\textwidth][c]{
\resizebox{1.05\textwidth}{!}{

\begin{tabularx}{\linewidth}{|Y|Y|Y||Y|Y|Y|Y|Y|Y||Y|Y|Y|Y|Y|Y||Y|}
\hline

\textbf{Ref.} &\textbf{Photo\-active material}&\textbf{Neuro functi\-onality}& \textbf{Neur\-onal Model} & \textbf{Metho\-dology (Device /Simulation)}& \textbf{Spike Rate (Hz)}& \textbf{Spike Width (\textmu s)}& \textbf{TFS (ms)}& \textbf{EPS (pJ)}&\multicolumn{6}{c||}{\textbf{Retinal Preprocessing Functionality}} &\textbf{Appli\-cation}\\
\hline

& & & & & & & & & WIS & SFA &\textbf{VA}  &  \multicolumn{2}{c|}{\textbf{CSRF}}&\textbf{CSRF\-+VA}&\textbf{Classif\-ication} \\
\hline
\hline
& & & & & & & & &&&  &\textbf{Lumin\-ance contrast}& \textbf{Color Opponency}&&\\
\hline
Biolo\-gical retina \ \cite{Laswick2024, Berry1997, OBrien2002, Tenglics2019, TTFSLi2024, Masland2017} & opsin & Photo-trans-ducer & close to HH & PC, BC, RGC, HC, and AC & \makecell{0.5 -\\ 500} & \makecell{0.6k -\\ 3k} & \makecell{30 -\\ 240}& $-$ & \multicolumn{6}{c||}{\cellcolor[HTML]{B3F0DF} Yes} &\cellcolor[HTML]{B3F0DF}Yes \\
\hline

Nat. Com. (2021) \ \cite{Radhakrishnan2021} & Si & Photo-trans-ducer& $-$& Si PD + MoS$_2$ FET&100& $-$ & $-$ &5 (White Light) &\cellcolor[HTML]{B3F0DF}Inten\-sity encoding & No& No &No&No&No& \cellcolor[HTML]{FFC7C5}C-SNN on MNIST \\
\hline

Nat. Com. (2023) \ \cite{Wang2023} & b-AsP/ MoTe$_2$ & Photo-encoder& $-$& hetero\-stru-cture FET& 100k & $-$ & $-$ & $-$ & No & No & \cellcolor[HTML]{B3F0DF} Yes & No & No &No& \cellcolor[HTML]{FFC7C5}C-SNN on MNIST \\
\hline

Nat. Com. (2024) \ \cite{TTFSLi2024} & In$_2$O$_3$ & Syna-ptic PT & $-$& In$_2$O$_3$ PT + NbO$_x$ MR& 0.35M - 1.85M& $-$ &  (1.04 - 13.00)$\times$ $10^{-3}$ &1.06 $\times$ $10^{-3}$ (365 nm) &\cellcolor[HTML]{B3F0DF}Inten\-sity encoding & No& No &No&No&No& \cellcolor[HTML]{FFC7C5}C-SNN on ResNet-18 for self-driving car\\
\hline

ACS Nano (2021) \ \cite{Pei2021} & PbS QDs & Optical Synapse & LIF & photo\-electric MR + MoS$_2$- MoO$_x$ TS MR & 160 & $-$ & $-$ & $-$ & No, optical spiking & \cellcolor[HTML]{B3F0DF}Yes & No & No & No &No& Auto\-mobile meeting (MATLAB)\\
\hline

Adv. Fun. Mat. (2024) \ \cite{LiH2024} & MoS$_2$ & Optical neuron & LIF & 1T-1C PT & $-$ & $-$ & $-$ & $-$ & No, optical spiking &No & No & No & No &No& No \\
\hline

Adv. Mat. Tech. (2025) \ \cite{Aung2025}  & MoS$_2$ & Optical neuron & LIF & PT & $-$ &$-$  & $-$ & $-$ & \cellcolor[HTML]{B3F0DF}Wavel\-ength encoding&No & No & No &No &No& \cellcolor[HTML]{FFC7C5}C-SNN on CIFAR \ 10 and DVS128\\
\hline

Adv. Mat. (2025) \ \cite{Kim2025} & InGa\-ZnO & Opto-MR & IF & Opto-MR + Si neuristor & 185 - 600 & 2k - 5k & $-$ & 30 (White Light) & \cellcolor[HTML]{B3F0DF}Inten\-sity encoding & No & No & \cellcolor[HTML]{B3F0DF}Yes for RGCs & No &No& \cellcolor[HTML]{FFE3E2}RF+ C-SNN for finger\-print classification \\
\hline

Nat. Com. (2024) \ \cite{Laswick2024} & Si PDs & Photo-trans-ducer & HH & PDs + 2L (p/n) organic-T (Na/K-Ch, HC) + LTSpice & 0.5 - 25 & $-$ & $-$ & 1.1 $\times$ $10^{6}$ (580 nm)& \cellcolor[HTML]{B3F0DF} yes & No & No & No & No &No& No \\
\hline
\hline

This Work& WSe$_2$ & Light-modul\-ated Na Ch or PC & HH & 2DSpT, Cadence Virtuoso (Spectre) & 0 - 2k & 11 - 323 & 4.2$\times$ $10^{-3}$ - 1.25 & 0.9 (D), 2 (480 nm), 24.5 (800 nm) & \cellcolor[HTML]{B3F0DF}yes & \cellcolor[HTML]{B3F0DF}yes &  \cellcolor[HTML]{B3F0DF}yes & \cellcolor[HTML]{B3F0DF}yes & \cellcolor[HTML]{B3F0DF}Yes&\cellcolor[HTML]{B3F0DF}Yes &\cellcolor[HTML]{B3F0DF} PR+ CSRF+ VA+ BE+ C-SNN on F- MNIST and camou\-flaged datasets (COD-10K, synthetic) \\

\hline

\end{tabularx}
}}
\vspace{2cm}
{\footnotesize $*$Photodiode (PD); Phototransistor (PT); Memristor (MR); Conventional SNN (C-SNN)}
\end{center}

\newpage

\bibliography{mainref}